# Antipolar and short and long-range magnetic ordering in quasi-two-dimensional AgCrP$_2$S$_6$

Chaitanya B. Auti[1, *], Atul G. Chakkar[1], Shantanu Semwal [2], Sebastian Selter[3], Yuliia Shemerliuk[3], Bernd Büchner[3,4], Saicharan Aswartham[3], Koushik Pal[2], and Pradeep Kumar[1#]

[1]*School of Physical Sciences, Indian Institute of Technology, Mandi-175005, India.*

[2]*Department of Physics, Indian Institute of Technology Kanpur, Kanpur 208016, India*

[3]*Institute for Solid State Research, Leibniz IFW Dresden, Helmholtzstr. 20, 01069 Dresden, Germany*

[4]*Institute of Solid State and Materials Physics and Würzburg-Dresden Cluster of Excellence ct.qmat, Technische Universität Dresden, 01062 Dresden, Germany*

**Abstract**

Within the Landau theoretical framework, the decreased entropy with decreasing the temperature is accompanied by the symmetry breaking and hence a corresponding phase transition. The broken symmetries leave its imprint on the underlying excitations and the same may be gauged using renormalization of these excitations. AgCrP$_2$S$_6$ provides a versatile playground to probe dynamics of the quasiparticle excitations as well as multiple phase transitions with lowering temperature linked with the polar, lattice and spin degrees of freedom. Here, we report an in-depth temperature- and polarization-dependent Raman scattering measurements on single crystals of quasi 2D zigzag antiferromagnet AgCrP$_2$S$_6$ along with the first principle based phonon calculations. We observed multiple phase transitions triggered by the short and long-range ordering of spins at ~ 90 K and 20 K, respectively; within the Cr sublattice where spins are arranged in a 1D chain, evident by the distinct anomalies in the phonon modes self-energy parameters as well as intensity. Contrary to the conventional belief, we uncovered potential quasi-antipolar ordering at ~ 200 K and with further lowering in temperature an antipolar ordering at ~ 140 K attributed to the Ag ions, which is conjectured to be forbidden owing to the heaviness of Ag ions. The quasi-antipolar and antipolar ordering is gauged via the distinct renormalization of the phonon's parameters, which survives at all the temperatures. Additionally, large number of modes appears with decreasing the temperature, in the window of ~ 200-140 K, where antipolar ordering starts settling in. The emergence of large number of phonon modes below ~ 200 K, nearly double of those at room temperature, suggests the lowering of symmetry from high temperature $C_{2h}$ to the low temperature $C_2$ or $C_s$ and as a result doubling of the unit cell.

*autichaitanya5@gmail.com

#pkumar@iitmandi.ac.in

## 1. Introduction

Two-dimensional (2D) van der Waals (vdWs) materials have attracted significant attention due to their distinct properties that are advantageous for potential applications in next-generation electronics as well as rich physics [1,2]. A significant domain of modern research focuses on magnetic vdWs materials that display intriguing physical properties. The intrinsic 2D nature of these materials opens up unique opportunities to explore novel magnetic ground states, magnetic excitations, and their interactions with other quasiparticle excitations [3-6]. These tunable properties are of great interest not only for fundamental studies but also for potential applications in advanced fields such as spintronics, optoelectronics, and nanoelectronics [7,8].

Recently, 2D metal thiophosphates (MTPs) have drawn significant interest because of their intriguing characteristics, including magnetism down to monolayer, ferroelectricity, and multiferroicity. The $MPS_3$ (M = Mn, Fe, Ni, Co, V) family of layered materials presents a rich array of isostructural compounds that form 2D vdWs antiferromagnets [6,9,10]. 2D $MPS_3$ contain a wider range of band gap of ~ 0.24-3.5 eV, making them highly attractive for strategic applications [11]. Colombet et al. reported that heterocharge substitution also yields numerous stable compounds, which provide an extra degree of freedom in the atomic arrangement [12,13]. One such member of this family is AgCrP$_2$S$_6$, which exhibits an antiferromagnetic transition temperature ($T_N$) at ~ 20 K [14]. The non-magnetic Ag atom and the magnetic Cr atom are arranged in zigzag strips. The Ag$^+$ ion has bigger size compared to the Cr$^{3+}$ ion; as a result, the layer above the Ag atoms significantly bends outward (see Supporting Information Fig. S1-(a)). Consequently, a deformation occurs in the P$_2$S$_6$ octahedra to accommodate the different sizes of the cations. Therefore, to reduce this strain energy, the cations organize themselves in a zigzag chain formation, this arrangement functions as a one-dimensional chain of magnetic atoms with a spin value of S = 3/2 (see Supporting Information Fig. S1-(d)). The layers are arranged in AAA stacking, resulting in a quasi-2D structure [15,16].

Magnetic fluctuations in 2Ds are influenced by the symmetry of the order parameters, namely the Ising, XY, and Heisenberg types [9]. One-dimensional (1D) spin S = 3/2 Heisenberg antiferromagnetic Hamiltonian is given as [17] :

$$H = -J\sum_i (S_i^x S_{i+1}^x + S_i^y S_{i+1}^y + \lambda S_i^z S_{i+1}^z) + J'\sum_{i,k} S_i S_k + D\sum_i (S_i^z)$$, where $\lambda$ and D denotes the exchange and single-ion anisotropy parameters, $S_i^x$, $S_i^y$ and $S_i^z$ are the x, y, and z components of the total spin at $i^{th}$ site. J and J' are the intra- and interchain exchange coupling constants, respectively. A very small value of | J'/J | ~ 10$^{-4}$ (for AgCrP$_2$S$_6$) confirm that the system comprises of 1D spin chains. The Cr-(P or S)-Cr angle of around 90$^0$ allows a strong direct overlap of the octahedrally surrounded Cr's half-filled t$_{2g}$ orbitals [17]. Mutka et al. [14] shows gapless excitation spectrum in AgCrP$_2$S$_6$ probed via inelastic neutron scattering and spin-wave velocity V = 4.6J, which is significantly higher than the classical estimate of 3J, suggesting that quantum effects dominate the behaviour of spin waves in this system. The susceptibility measurements also indicate that the spin chains in AgCrP$_2$S$_6$ exhibit 1D behaviour far above the $T_N$ (~ 20 K). Recent work by Park et al. [18] also suggests that the weakest interaction is

interchain, leading to essentially 1D magnetic behaviour in each layer, while two layers exhibit a weak ferromagnetic coupling. The anisotropic structure introduces an additional degree of freedom, offering the potential to manipulate the material's structure in different ways to provide tunability for probing fundamental magnetism with reduced dimensionality.

Raman spectroscopy is a powerful technique for studying the presence of magnons, phase transitions, and various quasiparticle excitations, including spin-phonon and electron-phonon coupling, in 2D magnetic materials [19-23]. Recently, heterocharge MTPs moved into focus due to their low-dimensional magnetism, multiferroicity, and Haldane phase [24,25]. Understanding these singular behaviour in MTPs will provide valuable insights for potential applications in spin-based logic gate devices, non-volatile memory storage devices, and the possibility of fabricating heterostructures for multifunctional device applications [26,27]. Motivated by the rich underlying physics in these low-dimensional systems, we conducted an in-depth investigation of the $AgCrP_2S_6$ single crystal using Raman scattering measurements as well as first principle calculations of the phonons. The absence of current literature on the temperature-dependent characteristics of the phonon and magnetic dynamics, suggesting a potential for uncovering novel insights into materials vibrational as well as magnetic properties.

In this work, we elucidate the phonon excitations and crystal symmetry of $AgCrP_2S_6$ using comprehensive temperature and polarization-dependent Raman spectroscopy measurements along with the first principle calculations of the phonons. We observed significant hardening of the low energy phonon modes, attributed to the strong phonon-phonon anharmonic interactions, or an increase in bonding stiffness, indicating structural instability in the material. Based on the temperature-dependent phonon modes analysis, we observed a signature of antiferromagnetic transition at $T_N \sim 20$ K. Interestingly, we observed anomalies in mode frequency, linewidth, and intensity around temperatures ~ 90 K, ~ 140 K, and ~ 200 K. These transitions are attributed to the possible short-range ordering of the spins and polar ordering in the system, respectively. We also examined the interactions between quasiparticle excitations and electronic and/or magnetic continuums, i.e., Fano asymmetry in the phonon line shape, as a function of temperature. Polarization-dependent measurements show the two-fold symmetry of all observed modes in line with the group theoretical predictions. Quite surprisingly we also observed a systematic rotation of the polarization major axis of the phonon modes as a function of temperature, which opens the possibility of the tunability of the scattered light in this quasi 2D magnet via symmetry control.

## 2. Experimental and computational details

### 2.1. Experimental details

The single crystal samples of $AgCrP_2S_6$ were prepared by using chemical vapor transport technique as described in reference [28]. Temperature-dependent Raman scattering and photoluminescence (PL) measurements (see Supportive Information for PL measurements) were carried out from 6 to 300 K with ±0.1 K precision (Montana CCR) and the LabRAM HR-Evolution Raman spectrometer in the backscattering configuration. A 633 nm laser was used to excite the Raman spectra. Laser power was kept very low (< 0.5 mW) to prevent local heating. The sample surface was focused using a long working distance objective with 50X (NA = 0.5) magnification. Scattered light was collected through a Peltier-cooled charge-coupled detector. Polarization dependent Raman measurements were performed using a half wave plate and an analyzer at four distinct temperatures, i.e., 6 K, 60 K, 120 K, and 170 K.

### 2.2. Computational details

We performed first-principles density functional theory (DFT) calculations using the Vienna Ab initio Simulation Package [29,30] and utilized pseudo potentials generated using the projected augmented wave (PAW) [31] method. The valence electron configurations used are $4p^6 4d^{10} 5s^1$, $3d^5 4s^1$, $3s^2 3p^3$, $3s^2 3p^4$ for Ag, Cr, P and S respectively. We used gaussian smearing for treating the partial occupancies for each orbital with energy broadening of 0.05 eV and treated the exchange-correlation energies of the electrons within the GGA Perdew-Burke-Ernzerhof (PBE) functional [32]. The cutoff for the kinetic energy was set to 450 eV for all calculations. The primitive cell structure of $AgCrP_2S_6$ contained 20 atoms (2 Ag, 2 Cr, 4 P and 12 S atoms), we considered the antiferromagnetic (AFM) arrangement of Cr atoms and used the DFT-D3 method of Grimme [33] to include the vdW interactions in the structure. The lattice parameters were fully optimized using a k-point grid of 9x9x9 and the energy convergence criteria was set to $10^{-8}$ eV. After full optimization the energies are converged and the force on each atom is of the order of $10^{-4}$ eV/A or less. The lattice parameters and volume for the optimized structure were a=5.888 A, b=10.647 A, c=6.704 A and 403.848 $A^3$, respectively. The changes of a, b, c and volume are 0.09%, 0.24%, -0.6% and 0.3%, respectively when compared with the experimental lattice parameters.

The phonon calculations were done using the finite-difference method as implemented in Phonopy [34,35]. We used a 2x2x2 supercell containing 160 atoms, considering AFM arrangement of Cr atoms and including the vdW interactions. The tabulated frequencies and symmetry labels of all the modes at $\Gamma$-point are provided in Table S2.

## 3. Results and discussion

### 3.1 Crystal structure and phonon modes

$AgCrP_2S_6$ exhibits monoclinic symmetry having space group *P2/a* (#13) and point group $C_{2h}$ (2/m). Unit cell of the $AgCrP_2S_6$ bulk single crystal consists of 60 phonon modes at the $\Gamma$-point of the Brillouin zone, with the following irreducible representations [36]: $\Gamma_{total} = 14A_g + 14A_u + 16B_g + 16B_u$. There are 30 Raman active ($\Gamma_{Raman} = 14A_g + 16B_g$) modes and 27 infrared active ($\Gamma_{IR} = 13A_u + 14B_u$) modes (for more details see Table I). The Raman spectra is recorded using 633 nm laser excitation, and 48/27 Raman modes are observed at 6 K/300 K, which are labelled as P1-P48 at 6 K, as shown in Fig. 1(a). To extract the self-energy parameters of phonon modes, i.e., peak frequency ($\omega$), full width at half maximum (FWHM), and intensity, spectra are fitted using a sum of Lorentzian functions. We also recorded spectra in the higher frequency range of 700-1400 cm$^{-1}$ and observed some very weak higher order modes, as shown in Supporting Information Fig. S7. Figure 1(b) shows the temperature evolution of the Raman spectrum of AgCrP$_2$S$_6$ in the temperature range 6 to 300 K from 15-700 cm$^{-1}$ (magnified spectra in the smaller spectral range is shown in Supporting Information Fig. S4, Fig. S5, and Fig. S6).

All higher frequency modes ranging from ~ 250 cm$^{-1}$ are softens with increasing temperature, showing normal anharmonic phonon behaviour; while some lower frequency range modes show anomalous hardening, and others show a softening and get broadened with increasing the temperature. In the M$_2$P$_2$S$_6$ material family, the higher frequency (above ~ 250 cm$^{-1}$) modes are assigned to the vibration due to (P$_2$S$_6$) octahedra; therefore, these modes are similar in all M$_2$P$_2$S$_6$ compounds, and the lower frequency (below ~ 250 cm$^{-1}$) Raman modes are assigned to the heavy metal ion [37]. The particular M atom determines the cutoff frequency that separates the high- and low-frequency regimes. Phonon dispersion along with the atom resolved phonon density of state (phdos) is shown in Fig. 2a and Fig. 2b, respectively. From the atom resolved phdos in Fig. 2b we can see that in the high frequency region (500 cm$^{-1}$ to 650 cm$^{-1}$), the contributions to the phonon dispersion are mainly due to the P and S atoms due to smaller masses of these atoms. In the middle region (150 cm$^{-1}$ to 500 cm$^{-1}$) region we see the P contribution waning and being taken over by the S atoms and rising contribution due to Cr seen around 350 cm$^{-1}$, 300 cm$^{-1}$, and at 200 cm$^{-1}$. Finally at the lower frequency region (20 cm$^{-1}$ to 150 cm$^{-1}$) we see the contribution of S, P and Cr decreasing and increasing Ag contributions with large Ag localization with peaks around 60 cm$^{-1}$ and 20 cm$^{-1}$ to 40 cm$^{-1}$. All prominent phonon modes intensity reduces drastically from the lowest recorded temperature to the highest temperature as shown in the inset of Fig. 1(b). The most intense phonon mode P19 could be a mode associated with the anionic complex $[P_2S_6]^{4-}$, which is almost twice as intense as the other prominent modes P11, P33, and P47. Many phonon modes as P4, P5, P9, P15, P16, P17, P20, P23, P25, P29, P20, P31, P32, P34, P35, P36, P40, P41, P42, P44, and P48 disappear with increasing the temperature, as shown in the highlighted regions in Fig. 1(b), also see Fig.2, which shows disappeared phonon modes. Mode P31 (~ 469.5 cm$^{-1}$ at 6 K) shows an asymmetric line shape (see Fig. 5(c)), discussed in detail in Sec. 3.4.

### 3.2 Temperature dependence of the phonon modes

Generally, mode frequencies are expected to exhibit softening with increasing temperature [38]. Figure 4 shows the temperature dependence of frequency and FWHM of some of the prominent phonon modes labelled as P2, P3, P5, P7, P8, and P10-P14. (For more phonon modes, see Supporting Information Fig. S2). Some intriguing features are observed: (i) For the lower frequency range $\omega \leq 250$ cm$^{-1}$, some of the phonon modes P2, P3, P7, P10, P12, and P14 show abnormal hardening with increasing the temperature. (ii) Rapid decrease of intensity of the phonon modes P10 and P14 with increasing the temperature (see Fig. 5(a)). (iii) Many phonon modes disappear with increasing the temperature (see Fig. 3). Interestingly, pairs of two consecutive phonon modes P7-P8, P10-P11, and P13-P14 show opposite behaviour in their frequencies, while mode P2-P3 shows similar behaviour. Mode P7 shows abnormal hardening with increasing the temperature and approaches towards P8 ($\omega_{P8}$ = 159.5 cm$^{-1}$ at room temperature). Mode P10 approaches towards P11 with a sharp increase of frequency $\Delta\omega_{P10}$ ~ 12 cm$^{-1}$. However, P13 and P14 diverge from each other with increasing temperature (see Fig. 5(b)). P2 and P3 both show abnormal hardening with increasing the temperature. These modes may be associated with heavy metal ion Ag$^+$. The possible reasons associated with the anomalous frequency change are discussed in detail in Sec. 3.3.

Interestingly, phonon mode P5, P6, P8, P12, P13, P14, P19, P21, and P47 exhibit anomalous softening in frequency below ~ 20 K, i.e., antiferromagnetic ordering $T_N$ (~ 20 K).

The FWHM of the modes P2, P3, P7, P8, P10, P13 also shows change in slope at ~ $T_N$. This anomalous nature of the phonon modes below $T_N$ may be attributed to the spin-phonon coupling. The shift in phonon frequency resulting from spin-phonon coupling ($\Delta\omega_{Sp-Ph}$) may arise due to the modulation of the exchange integral by the lattice vibrations and is given as: $\Delta\omega_{Sp-Ph}(T) = \omega(T) - \omega_{anh}(T) = \lambda * \langle \vec{S_i}.\vec{S_j} \rangle$, where $\omega_{anh}(T)$ is the frequency due to phonon anharmonicity. $\langle \vec{S_i}.\vec{S_j} \rangle$ is the spin-spin correlation function with $i$ and $j$ as site indices. $\lambda$ is the spin phonon coupling constant, which describes the coupling strength between spins and phonons, and its value can be positive or negative depending on the nature of phonon modes. It is noted that spin-phonon coupling is generally expected to be effective only up to long range ordering temperature where the long-range magnetic correlations exist [39]. We note that at room temperature phonon modes are broad (see inset of Fig. 1(b), Fig. S4-S6), which develops into very sharp modes at low temperature, suggesting the presence of cation disorders at high temperature. For many phonon modes, e.g. see modes P2, P3, P7, P11, P12 in Fig. 4, change in the linewidth from room temperature to the low temperature is as large as ~ 75% ($|(\omega_{6K} - \omega_{300K})|/\omega_{300K}$).

It is noteworthy that some of the phonon modes P5, P6, P8, P13, P19, P21, P24, P28, P33, P39, and P47 show the change in slope in frequency shift at ~ 90 K (see Fig. 4 and Fig. S2), Mode P3 shows sharp increase of frequency up to 90 K, after which it remains nearly constant with further increase in temperature. Additionally, FWHM of the modes P13, P28, P45, and P47 exhibits a change in slope at ~ 90 K, thus highlighting a clear anomaly at ~ 90 K. Mode P10 shows a small kink in the frequency at ~ 140 K, while modes P13, P19, and P21 shows discontinuity in the frequency shift at this temperature. The FWHM of the phonon modes P6, P10 and P14 shows a jump at ~ 140 K, while that of modes P8, P19, and P47 shows a change in slope, suggesting some underlying anomaly at this temperature. Mode P2 shows a discontinuity in frequency at ~ 200 K, while mode P8, P19, P21, and P39 shows a change in slope in frequency shift at this temperature. Mode P12 shows an interesting frequency behaviour for temperature above 200 K, it exhibits a hardening from ~ 204.5 cm$^{-1}$ to ~ 209 cm$^{-1}$ with increasing temperature from 200 K till ~ 300 K, showing a clear anomaly in the frequency around ~ 200 K. FWHM of mode P6 and P10 shows a jump at ~ 200 K; while mode P7, P12, P19, and P21 shows change in slope, showing anomaly at ~ 200 K. This clearly shows that the lifetime of the phonons is strongly influenced in the vicinity of this temperature. Also, many phonon modes start appearing below ~ 200 K and continue to emerge till ~ 140 K (see Fig. 3) consistent with the anomalies observed in the modes which survive at all the temperatures.

Interestingly, intensity change of all phonon modes with temperature is nearly uniform except for the modes P10 and P14 whose intensity decreases sharply with increasing temperature by ~ 10 times compared to its intensity at 6 K, as shown in Fig. 5(a). Phonon modes P2 and P3 show very interesting intensity variation; their intensity increases with increasing temperature up to ~ 90 K, after which it decreases, indicating an anomaly at ~ 90 K. We note that similar intensity behaviour is also reported for the low-frequency modes in a sister compound i.e. $CuCrP_2S_6$, which suggest that lattice dynamics of off-centre $Cu^+$ cations within the intralayer are not quenched, indicating order-disorder-type of structural transition is responsible for the emergence of defect dipoles associated with Cu ions [40]. It is observed that the intensity of the modes P7, P10, P11, P13, P14 and P19 shows a sharp decrease with increasing temperature and remains nearly constant after 200 K clearly shows an anomaly at ~ 200 K.

Origin of the observed anomalies in the frequency, linewidth, and intensity at ~ 90 K, 140 K, 200 K is unresolved. However, the anomaly at ~ 90 K may be attributed to short range-spin ordering. A similar transition has also been suggested in other similar system, $AgCrP_2Se_6$, and it was argued that a local ferromagnetic order exists above $T_N$. The entropy change across $T_N$ is only ~ 56%, i.e. ~ 44% is already released above $T_N$ hinting development of short-range ordering much above $T_N$ [41]. Also, the exchange parameter for $AgCrP_2S_6$ is estimated to be of the order of J ~ 9meV (~ 100 K) [14]; therefore, the observed changes in the phonon modes around 90 K may be associated with the short-range ordering. We note that its sister compound, i.e. $CuCrP_2S_6$, undergoes antiferroelectric and antiferromagnetic ordering attributed to the $Cu^+$ and $Cr^{3+}$ cations, respectively. It undergoes a phase transition from a room temperature paraelectric state (space group *C2/c* and point group $C_{2h}$) to a quasi-antipolar state at ~ 190 K (space group $P_c$ and point group $C_s$) and with further lowering the temperature it goes to a fully antiferroelectric phase at ~ 145 K with space group $P_c$ (point group $C_s$) attributed due to the ordering of Cu ions within the lamellae. Additionally, across the antipolar phase transition at ~ 145 K it shows negative thermal expansion [20]. On the other hand, for $AgCrP_2S_6$ it is advocated that since $Ag^+$ is a heavier cation, it will not dislocate in the lattice, and hence no ferroelectric transition is expected [42]. Also, for $AgCrP_2Se_6$, two higher temperature transitions have been reported at ~ 68 K and 125 K, other than the low temperature antiferromagnetic transition at ~ 42 K [41]. The two transitions are attributed to the possible ferroelectric/antiferroelectric ordering due to lattice restructuring involving Ag ions. Also, a negative thermal expansion is reported around ~ 180 K. We note that Cu ion hopping is responsible for the ionic conductivity in $CuCrP_2S_6$. Ionic conductivity in $AgCrP_2S_6$ is low compared to the $CuCrP_2S_6$ as also reflected from their activation energy (1.26/0.67eV for (Ag/Cu)CrPS) [42]. The hopping of Ag ion in $AgCrP_2S_6$ cannot be ruled out, though the hopping may be limited and hence the possibility of an antipolar ordering similar to $CuCrP_2S_6$. Our observations of soft phonon modes (discussed in detailed in Sec. 3.3) and observed anomalies in the phonon modes clearly suggest that, despite the heaviness of $Ag^+$ ion, the antipolar and quasi-antipolar phase transitions cannot be ruled out in AgCrP$_2$S$_6$. Our observations suggest several contributing factors for this anomalous trend of the phonon modes, as discussed in Sec. 3.3. One possible explanation is that local structural contraction may be associated with the Ag$^+$ and Cr$^{3+}$ cations structures. Nevertheless, our observations of distinct changes in modes frequency, linewidth, and intensity suggest the possibility of multiple phase transitions in the AgCrP$_2$S$_6$. So far, no transport measurements have been done on this material to confirm a ferroelectric transition. Our observation calls for further experimental as well as theoretical studies to explore the complex physics associated with this system, particularly the ambiguity surrounding the ferroelectric phase transition.

Irreducible phonon mode calculations show that there are only 30 Raman active modes, but we have observed a total of 48 phonon modes at the lowest temperature, including some weak modes (see Table S1). As the temperature rises, several modes disappear, with the total number reducing to nearly half, i.e. 27, of that observed at low temperature, as shown in Fig. 3. While decreasing the temperature we notice many new modes starts appearing in the temperature window of ~ 140-200 K, as expected for a first-order character of the antiferroelectric transition. We note that with increasing temperature from 6 K, modes involving dynamics of metal ions soften and broaden and finally disappear around 140-200 K. For example, low frequency mode P4 /P5, see Fig. 4/Fig. S2, soften by ~ 2.5/3% and disappear at ~ 140/200 K. The disappearance of these phonon modes along with many other modes with increasing temperature can be attributed to several factors, including thermal broadening, increased anharmonic phonon-phonon interactions, and reduced phonon lifetime, which results in an increase in FWHM. Additionally, structural instability in symmetry may also contribute to the suppression of certain modes at elevated temperatures [43]. It is likely that similar to its sister compound CuCrP$_2$S$_6$, AgCrP$_2$S$_6$ also undergoes order-disorder type of structural transition and is also responsible for the potential polar ordering. Our observations clearly favour the possibility of structural instability, with singular changes in self-energy parameters of the prominent modes which survive at all the temperatures. We note that symmetry of the newly emerged modes at low temperature (see Fig. S10, S11) is also similar to the modes at high temperature. This reflects that symmetry of the modes is same, but their number becomes nearly double i.e. going to the lower symmetry ($C_2$ or $C_s$) as temperature is decreased and doubling the unit cell hence giving double the number of modes than at room temperature. Our observation of emergence of new modes in the temperature interval of ~ 140-200 K and modes showing distinct anomalies at these temperatures is clearly indicative of a symmetry-lowering structural transition from a high temperature $C_{2h}$ point group to $C_2$ or $C_s$ point groups at these temperatures.

### 3.3 Discussion on anomalous hardening of the phonon modes

Anomalous hardening of the low frequency phonon modes P2, P3, P7, P10, P12 and P14 may be associated with the anharmonic effects which comes into the picture due to (i) Strong phonon-phonon anharmonic interactions and (ii) Lattice thermal contraction or structural instability, which are discussed as following:

**(i) Strong phonon-phonon anharmonic interactions**

Generally, with increasing temperature a phonon mode softens. However, a strong phonon-phonon anharmonic interactions may lead to phonon hardening with increasing temperature [44] and this abnormal phonon hardening in our case may also favour a possible transition based on the large atomic displacement i.e. antiferroelectric transition. These interactions play a crucial role in determining the ground state of incipient ferroelectrics. The self-energy of a phonon is given as: $\Sigma = \Sigma_r + i\Sigma_i$, imaginary part ($\Sigma_i$) is related to the lifetime of a phonon and for a given phonon, it is a function of $\omega$ and $T$ given as: $\Delta\Gamma(T) \propto \Sigma_i(\omega) \propto |V|^2 \rho_2(\omega)$, where $V$ is the average anharmonic coupling constant and $\rho_2(\omega)$ is the two-phonon density of states with the restrictions of momentum conservation and $\omega = \omega_1 + \omega_2$. The real part, $\Sigma_r$, is associated with the phonon frequency given as:

$$\Delta\omega(T) \propto \Sigma_r = \frac{2|V|^2}{\pi} \int_0^\infty \frac{\rho_2(\omega')}{\omega^2 - \omega'^2} \, d\omega', \tag{1}$$

for simplicity, $V$ is assumed as independent of frequency. The sign of this integral determines the sign of the self-energy constant (A), which is generally negative due to the dominance of the term with $\omega' > \omega$. Conversely, if the peak in the two-phonon density of states is lower than $\omega$, the integral may give a positive value, potentially resulting in a hardening of the phonon modes [44,45]. Our observation of the anomalous hardening of the mode P2, P3, P7, P10, P12, and P14 clearly reflects strong phonon-phonon anharmonic interaction suggesting large movement of ions. We note that these low frequency modes are associated with the motion of heavy ions $Ag^+/Cr^{3+}$, and these ions are the ones where displacement is potentially responsible for the antiferroelectric transition in this class of material. Therefore, based on the anomalies observed here we also suggest that these anomalies in this system are potentially related with the potential antipolar ordering.

Now to understand the impact of anharmonicity on the frequency and FWHM of the phonon modes in the temperature range of $\sim$ 140 to 300 K, we use anharmonic three-phonon model, given as [38,46]:

$$\Delta\omega = \omega(T) - \omega_0 = A\left(1 + \frac{2}{e^x - 1}\right), \tag{2}$$

$$\Delta\Gamma = \Gamma(T) - \Gamma_0 = C\left(1 + \frac{2}{e^x - 1}\right), \tag{3}$$

where $\omega_0$ and $\Gamma_0$ represent the mode frequency and FWHM at absolute zero temperature, respectively, and $x = \frac{\hbar\omega_0}{2k_B T}$. A and C represent the self-energy constant that describe the strength of phonon-phonon interactions involving three phonon processes. Anharmonic model fit to the experimental data for the case of frequency and FWHM is shown by solid red lines in the temperature range of 140-300 K. Figure 4 illustrates prominent modes fitted (frequency and FWHM) using equations (2) and (3), respectively. The optimal parameters are shown in Table S3. The self-energy constant (A) associated with mode P2 and P10 is positive, indicating that term $\omega > \omega'$ is dominant in equation (1) for these modes, while for all other phonon modes it is negative.

**(ii) Lattice thermal contraction or structural instability**

Temperature-dependent behaviour of the phonon modes is closely related to the direction (symmetry) and magnitude of lattice vibrations; anomalous phonon mode behaviour may arise from their anisotropic structural properties. Hardening of the low frequency phonon modes with increasing temperature indicates that bonding stiffness corresponding to these modes increases, suggesting that there may be a local structural instability [47]. This type of behaviour of phonon modes is referred to as soft phonon modes, i.e., their frequency decreases with decreasing temperature. The soft mode like behaviour is understood within the mean field approximation [40] $\omega(T) = \beta(T^* - T)^{1/2} + \omega_0$, where $\beta$ is constant and $T^*$ is the characteristic temperature at which a mode disappears. As clear from our observation $T^*$ is found to be in the range of ~ 140-200 K. We note that frequency of these modes does not go to zero across $T^*$, unlike the soft phonon mode expected in a displacive-type structural transition. The presence of second term $\omega_0$ favours the view point that quasi-antipolar order is also present in this system similar to compound $CuCrP_2S_6$ and is driven by an order-disorder transition rather than the condensation of a soft phonon mode. Many of the phonon modes (P2, P3, P7, P10 and P14; see Fig. 4) associated with motion of metal cations shows anomalous softening from the expected mean-field behaviour may be due to a renormalization of the phonon energy by interlayer Ag ion hopping motions. We note that similar behaviour is also reported for a phonon mode in CuCrP$_2$S$_6$ [40]. Therefore, our observations of anomalous softening of the low frequency modes in AgCrP$_2$S$_6$ clearly suggest the potential quasi-antiferroelectric and subsequently antiferroelectric ordering in the temperature window of ~ 200-140 K.

**3.4 Fano resonance**

We observed a phonon mode, P31, which shows an asymmetric line shape, and this may result due to the interaction between the phonon and underlying magnetic or electronic continuum, and this asymmetry may be analysed using the Breit-Wigner-Fano (BWF) function as [48]: $I(\omega) = I_0 \frac{[1+(\omega-\omega_0)/q\Gamma]^2}{1+[(\omega-\omega_0)/\Gamma]^2}$, where $\omega_0$ and $\Gamma$ are the frequency and FWHM of uncoupled phonon, respectively. $q$ is the asymmetry parameter that characterizes the coupling strength between the phonon and the continuum, quantified by the parameter $1/|q|$. Microscopically $1/|q| \propto |V_E|$, where $|V_E|$ tells the interaction between the discrete state and the continuum. Therefore, limit $1/|q| \to 0$ corresponds to weak coupling, which causes the peak to be symmetric and reduces to a Lorentzian line shape; and limit $1/|q| \to \infty$ represents strong coupling that causes the peak to be more asymmetric. Figure 5(c-i) shows the temperature evolution of asymmetric mode P31 in the temperature range of 6-300 K. Figures 5(c-ii) and (c-iii) show the temperature dependence of the frequency and FWHM of mode P31 in the temperature range of 6-200 K. For mode P31, as the temperature increases, the frequency decreases and the FWHM increases, with a change in slope for both parameters occurring around $T_N$ (~ 20 K). Figure 5(c-iv) shows the temperature dependence of the coupling strength (1/|q|), and it decreases with increase in temperature with a jump around $T_N$ (~ 20 K). This suggests the coupling between phonon and underlying magnetic continuum is strongly affected by long-range magnetic ordering.

**4. Polarization dependence of the phonon modes**

To determine the symmetry and to understand the angle-dependent characteristics of the phonon modes, we carried out an in-depth polarization-dependent Raman measurements at four temperatures: 6 K, 60 K, 120 K, and 170 K as shown in Fig. 6 for few selected modes (for more phonon modes see Supporting Information Fig. S8 and Fig. S9). The polarization-dependent measurements were performed by gradually rotating the incident light polarization using a half-wave plate (λ/2 plate) with increments of 20 degree, ranging from 0 to 180 degrees, while the polarization of the scattered light was fixed by using an analyzer. For convenience, we have mirrored the data points from 0 to 180 degrees. The angular dependence of the phonon modes intensity is given by eqn (4) and (5), considering the complex element of the Raman tensor (see Sec. S1 in Supporting Information for polarization intensity analysis details)

$$I_{A_g} = |b|^2 \cos^2(\theta + \theta_0)\cos^2\theta_0 + |c|^2 \sin^2(\theta + \theta_0)\sin^2\theta_0 \\ + 2|b||c|\cos(\theta + \theta_0)\cos\theta_0 \sin(\theta + \theta_0)\sin\theta_0 \cos\phi_{bc} \quad (4)$$

$$I_{B_g} = |f|^2 \sin(\theta + 2\theta_0), \quad (5)$$

where $\phi_{bc} = \phi_b - \phi_c$ is the phase difference between b and c components of the Raman tensor.

In Fig. 6 (Supporting Information Fig. S8 and Fig. S9) solid red line shows the fitted curves using eqn (4) and (5).

Figure 6 shows the polar plot for the intensity of the modes P7, P8, P13, P19, and P47 at four different temperatures. Mode P7 shows the quasi-isotropic symmetry at 6 K, and it approaches two-fold symmetry at 60 K; and it shows complete two-fold symmetry at higher temperatures 120 and 170 K with maxima around 20 and 200 degrees. Mode P8 shows the two-fold symmetry at all four temperatures, with maxima around 40 and 220 degrees. Mode P47 shows the quasi-isotropic symmetry at 6 and 60 K, while it becomes two-fold at 120 and 170 K with maxima around 20 and 200 degrees. The modes that disappeared at higher temperatures also exhibit twofold symmetry, as shown in Supporting Information Fig. S10 and Fig. S11.

This reflects that symmetry of these modes is also $A_g/B_g$ or A/B consistent with the group theoretical arguments. Interestingly, mode P13 shows the quasi-isotropic symmetry with maxima around 40 and 220 degrees at 6 and 60 K, while at higher temperatures 120 and 170 K, it shows the maxima around 90 and 270 degrees. Also, mode P19 and P33 (see Fig. S9) shows maxima at ~ 140 and 320 degrees at 6 and 60 K, while at higher temperatures it shows maxima ~ 0 and 180 degrees. This significant rotation in the polar plots with temperature may be associated with the magneto-optic Kerr effect, structural deformation, or changes in crystal symmetry. We note that similar rotation has also been reported recently in other 2D magnetic systems [5,49]. The microscopic understanding of these rotations and change in the symmetry from quasi-isotropic to twofold symmetry as a function of temperature calls for a detailed theoretical study to uncover underlying mechanisms responsible for these changes.

## 5. Conclusion

In conclusion, we performed a comprehensive temperature- and polarization-dependent Raman measurements on single crystal $AgCrP_2S_6$ along with the first-principle based calculations of the phonons. Our results have shown that this quasi 2D van der Waals material, $AgCrP_2S_6$, with 1D spin chains demonstrate a potential quasi-antipolar and antipolar phase transition opposite to the conventional view so far adopted in the literature, therefore putting forward that even the heavy ion (Ag) based system may show antiferroelectric ordering. Our findings also pave the way forward for potential controlling and manipulating these ordering as a function of layer and make them useful for potential applications. We also observed signature of a long-range magnetic ordering at low temperature, all these transitions are marked by the distinct renormalization of the phonon modes as well as emergence of multiple phonon modes with decreasing temperature. Our findings also point that this material may also be tuned as a potential multiferroic material similar to its sister compounds. Additionally, polarization-dependent measurements elucidate the role of anisotropy in the system, shedding light on the crystal symmetry. Our findings offer critical insights into the phonon dynamics of this quasi-2D magnetic compound, providing a deeper understanding of its underlying mechanisms.

**Acknowledgements**

P.K. thanks SERB (Project no. CRG/2023/002069) for the financial support and IIT Mandi for the experimental facilities.

**Figures:**

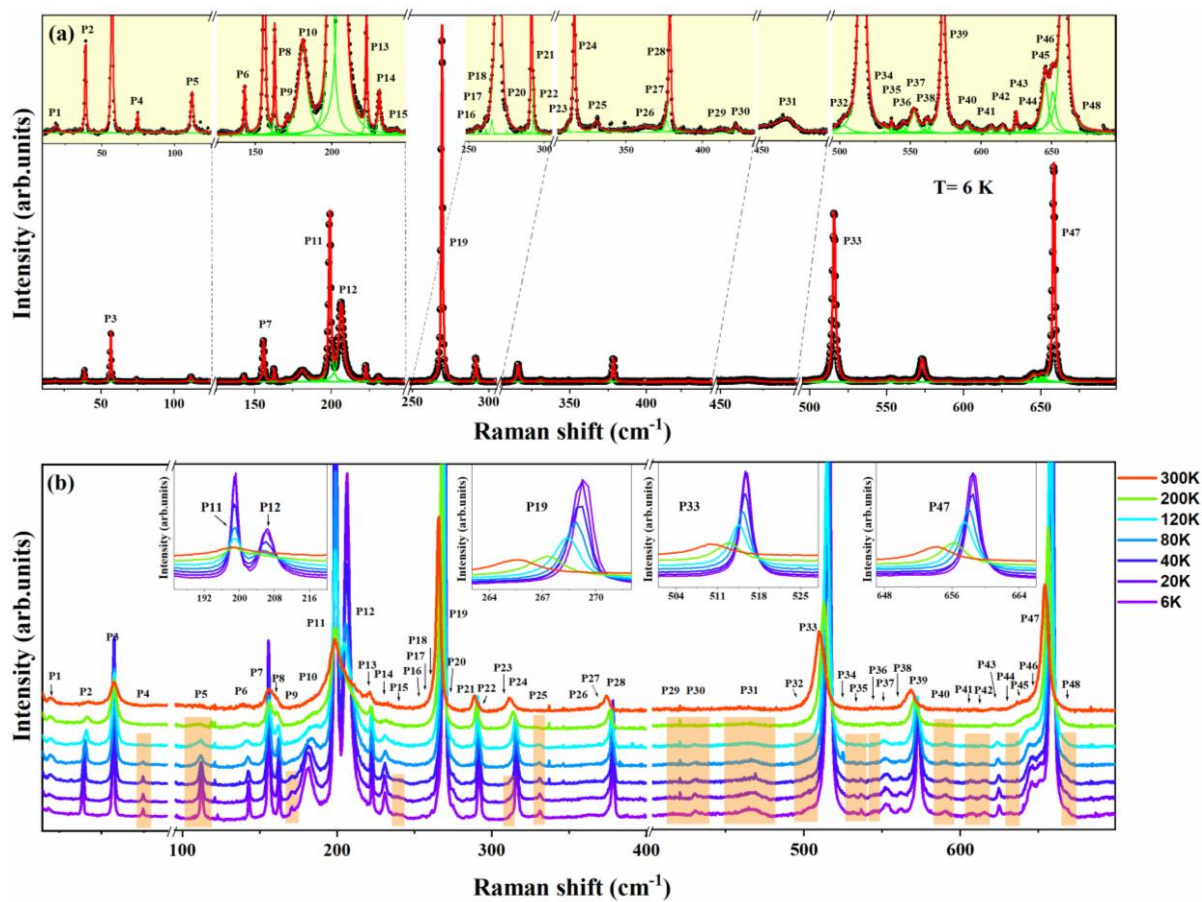

**Figure 1. (a)** Fitted Raman spectrum of the single crystal of AgCrP$_2$S$_6$ in the spectral range of 15-700 cm$^{-1}$ recorded at 6 K. The observed phonon modes are labelled as P1-P48. Solid thin line (green line) indicates the individual fit of the phonon modes and solid (red line) indicate the total fit of the Raman spectra. **(b)** Temperature evolution of the Raman spectrum in the temperature range of 6 to 300 K. Inset shows the magnified spectrum evolution of some prominent modes. Highlighted section in orange colour shows the disappeared modes with increasing temperature.

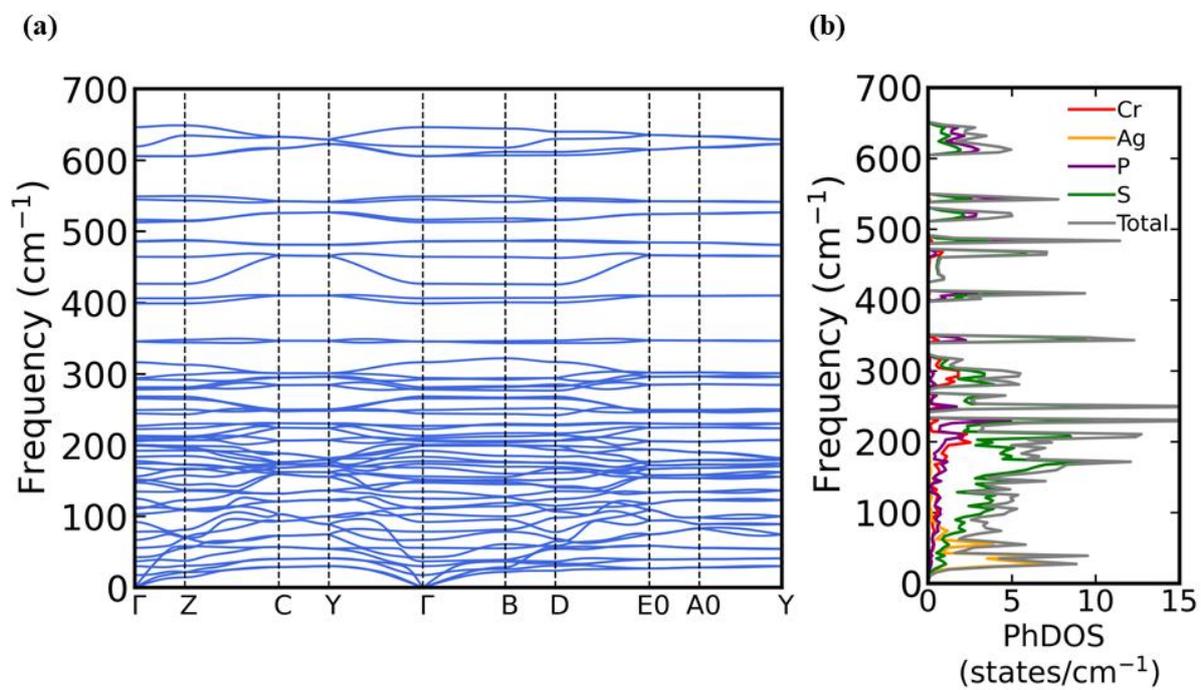

**Figure 2. (a)** Phonon dispersion and **(b)** atom resolved phonon density of states considering the AFM arrangement of AgCrP$_2$S$_6$ and including the vdW forces.

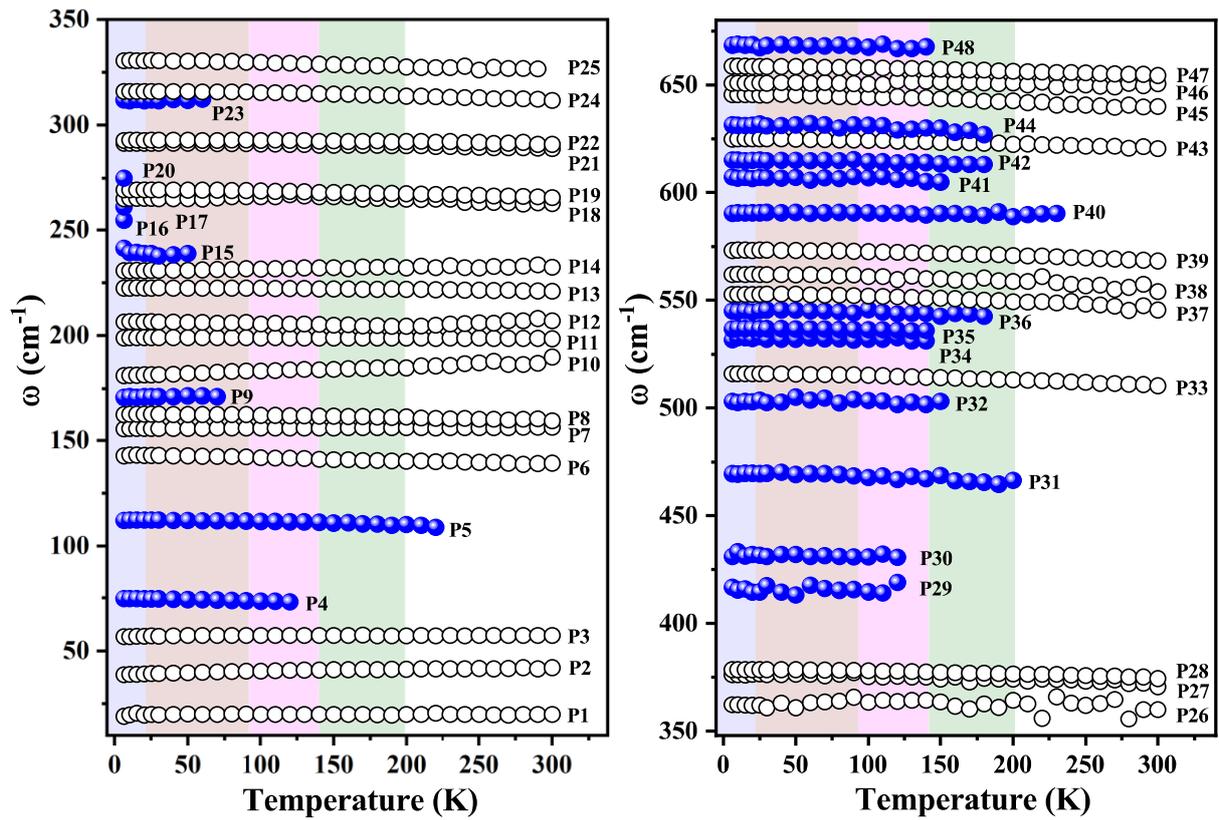

**Figure 3.** Temperature dependence frequency shift of all the observe phonon modes from P1-P48. Blue solid spheres represent the disappeared modes. Hollow circles represent phonon modes which survives at all temperatures. Blue, brown, pink, and green shaded regions show the temperature range of 0-20 K, 20-90 K, 90-140 K, and 140-200 K, respectively, which shows the regions of anomalies reflecting multiple transitions.

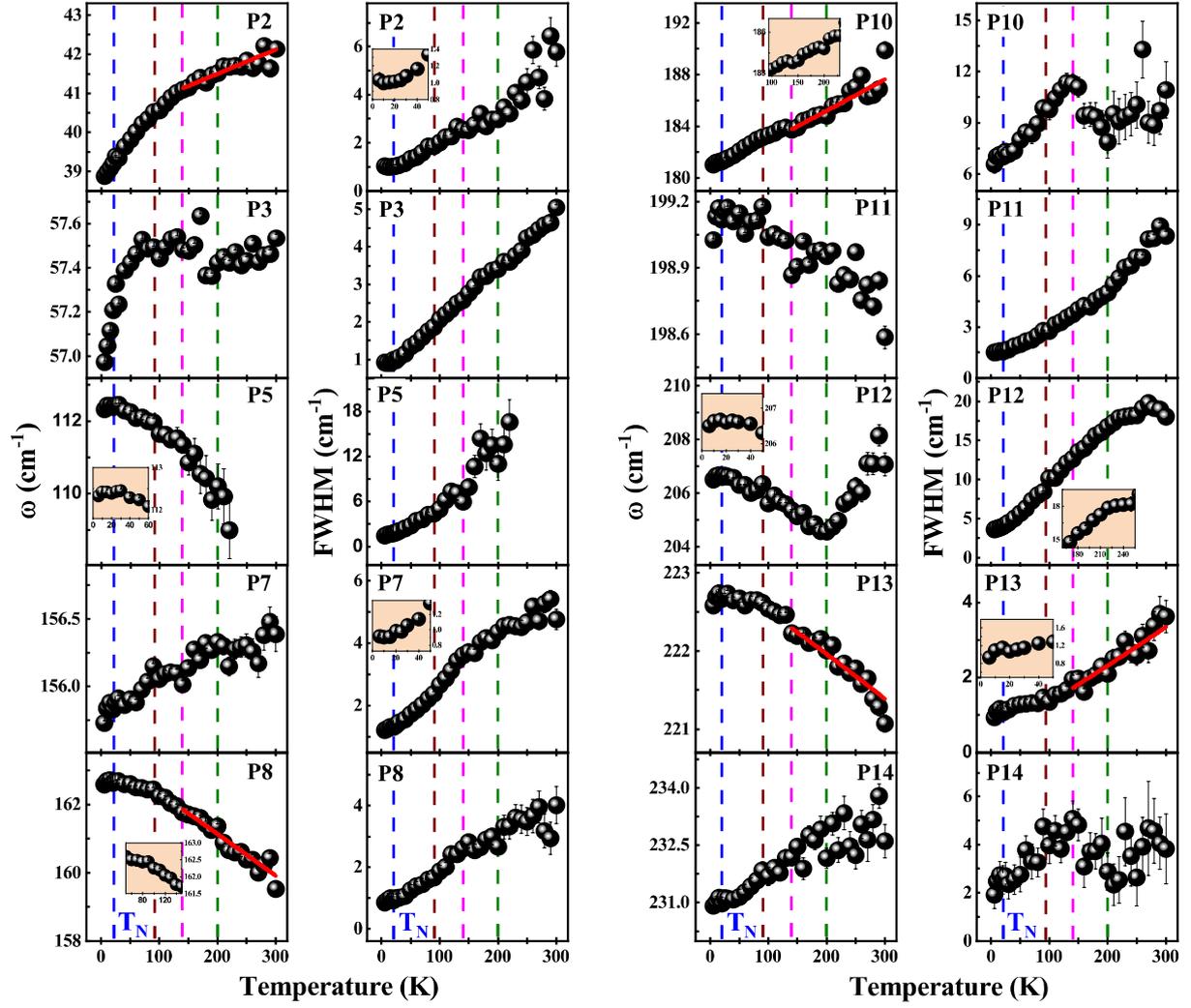

**Figure 4.** Temperature evolution of the mode frequency and FWHM of phonon modes P2, P3, P5, P7, P8, and P10-P14. The solid red line shows a three-phonon fitting in the temperature range 140 K to 300 K. The dashed blue line indicates the antiferromagnetic transition at $T_N \sim$ 20 K. The dashed brown, pink, and green lines reflect temperature anomalies at ~ 90 K, ~ 140 K, and ~ 200 K, respectively. Inset shows the magnified image of corresponding anomalies in frequency and FWHM.

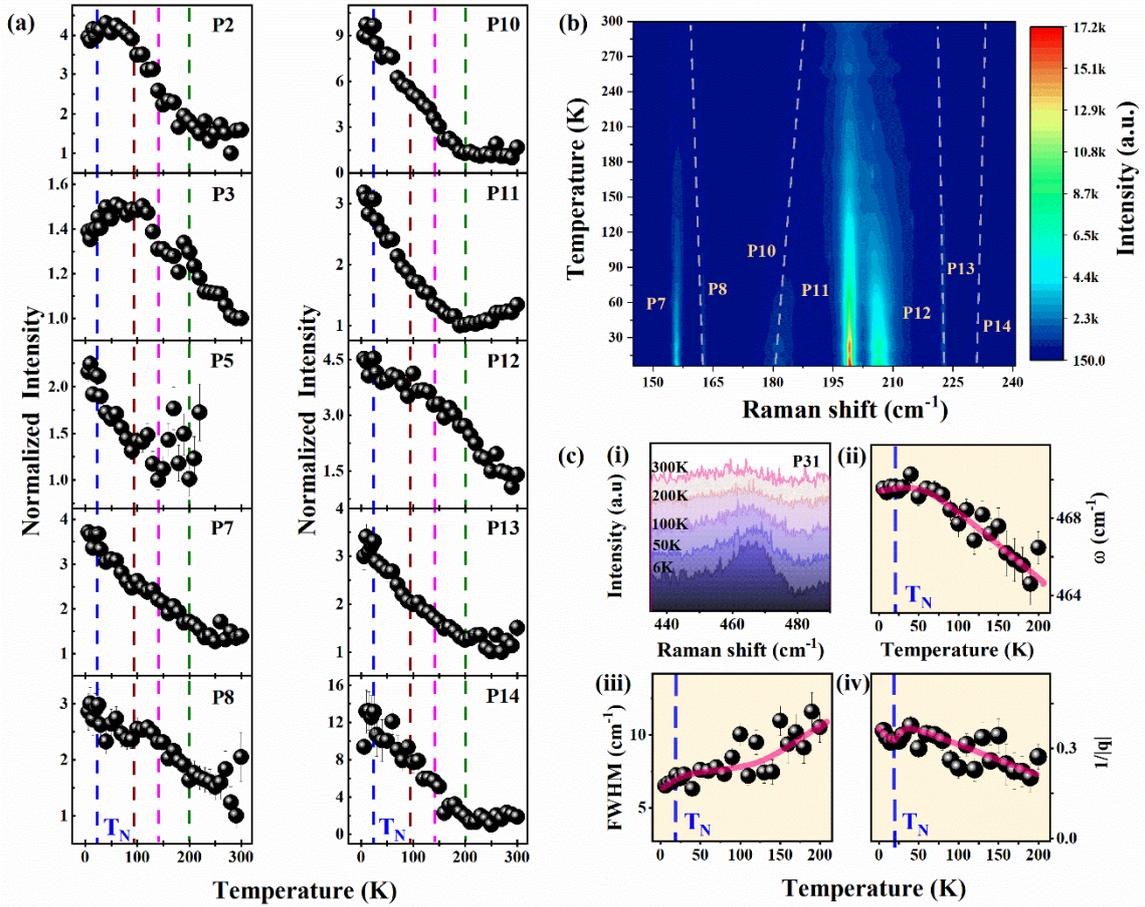

**Figure 5. (a)** Shows temperature dependence of the normalised intensity of modes P2, P3, P5, P7, P8, and P10-P14. **(b)** Shows a colour contour plot of the phonon intensity in the temperature-Raman-shift plane. The vertical dashed line marks the corresponding mode evolution. **(c)**-(i) Shows the temperature evolution of the Fano mode P31. (ii-iv) Shows the temperature dependence of the frequency, FWHM and the coupling strength (1/|q|) in the temperature range of 6 K to 200 K. The solid semi-transparent pink line is drawn as guide to the eye.

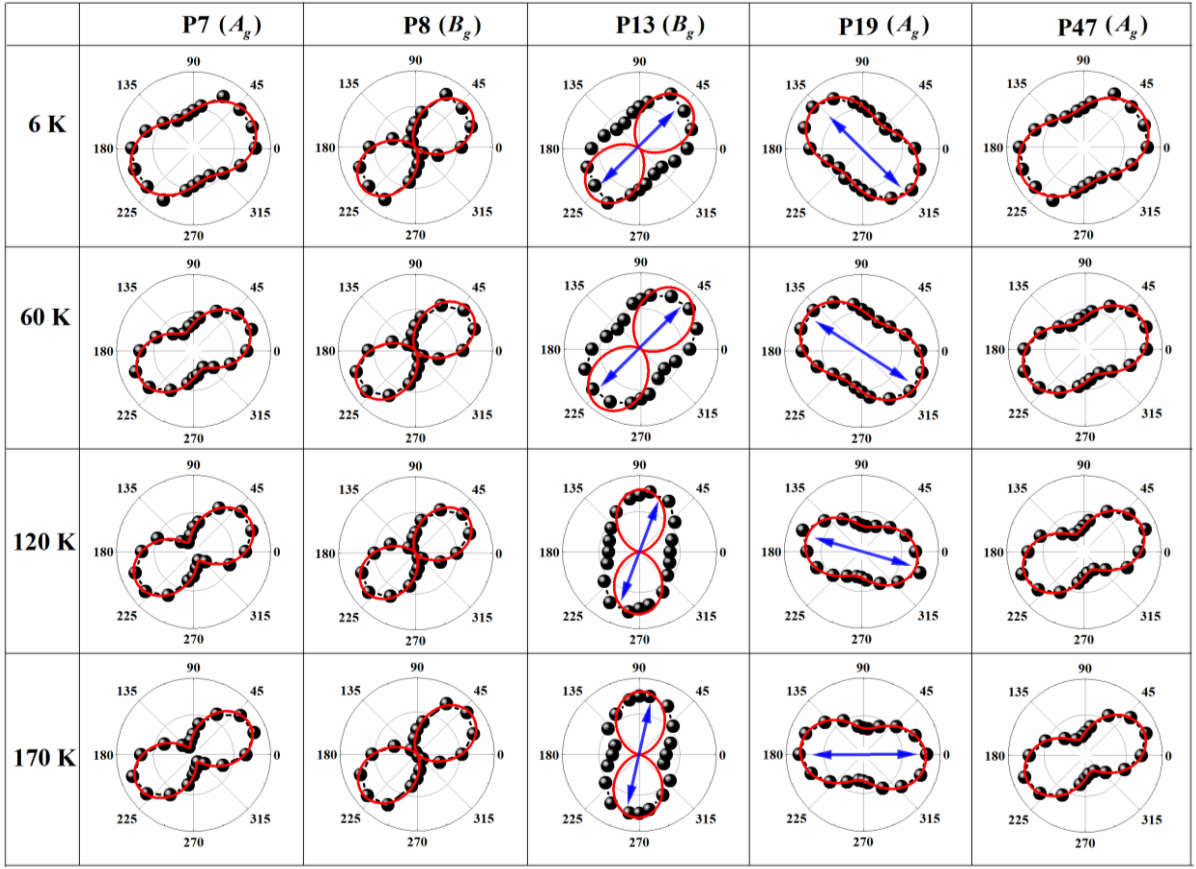

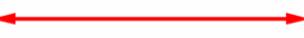

**Figure 6.** Polarization-dependent intensity of the phonon modes P7, P8, P13, P19, and P47 measured at temperatures of 6 K, 60 K, 120 K, and 170 K, respectively; with varying incident light polarization. Blue arrows represent the rotation of the axis, showing the maxima as a function of temperature. The solid red line at the bottom indicates the direction of the scattered light polarization.

**Tables:**

**Table I.**

Wyckoff positions of atoms and irreducible representations of the phonon modes for AgCrP$_2$S$_6$ in the monoclinic (P2/a; #13) crystal system.

| Atoms | Wyckoff position | Mode decomposition | Raman tensors |
|---|---|---|---|
| Ag | $2e$ | $A_g + A_u + 2B_g + 2B_u$ | $A_g = \begin{pmatrix} b & 0 & d \\ 0 & c & 0 \\ d & 0 & a \end{pmatrix}$ $B_g = \begin{pmatrix} 0 & f & 0 \\ f & 0 & e \\ 0 & e & 0 \end{pmatrix}$ |
| Cr | $2e$ | $A_g + A_u + 2B_g + 2B_u$ | |
| P | $4g$ | $3A_g + 3A_u + 3B_g + 3B_u$ | |
| S | $4g$ | $3A_g + 3A_u + 3B_g + 3B_u$ | |
| S | $4g$ | $3A_g + 3A_u + 3B_g + 3B_u$ | |
| S | $4g$ | $3A_g + 3A_u + 3B_g + 3B_u$ | |
| $\Gamma_{total} = 14A_g + 14A_u + 16B_g + 16B_u$; $\Gamma_{Raman} = 14A_g + 16B_g$; $\Gamma_{IR} = 13A_u + 14B_u$; $\Gamma_{Acoustic} = A_u + 2B_u$ | | | |

# Supporting Information:

## Antipolar and short and long-range magnetic ordering in quasi-two-dimensional AgCrP$_2$S$_6$

Chaitanya B. Auti[1, *], Atul G. Chakkar[1], Shantanu Semwal [2], Sebastian Selter[3], Yuliia Shemerliuk[3], Bernd Büchner[3,4], Saicharan Aswartham[3], Koushik Pal[2], and Pradeep Kumar[1#]

[1]School of Physical Sciences, Indian Institute of Technology, Mandi-175005, India.

[2]Department of Physics, Indian Institute of Technology Kanpur, Kanpur 208016, India

[3]Institute for Solid State Research, Leibniz IFW Dresden, Helmholtzstr. 20, 01069 Dresden, Germany

[4]Institute of Solid State and Materials Physics and Würzburg-Dresden Cluster of Excellence ct.qmat, Technische Universität Dresden, 01062 Dresden, Germany

## S1. Analysis of polarization dependent intensity:

The variation in intensity as a function of polar angle can be understood using a semi-classical approach. The Raman scattering intensity ($I$) is expressed as $I \alpha |\hat{e}_s^T . R . \hat{e}_i|^2$, where T is the transpose of the vector and R is the Raman tensor [1]. $\hat{e}_i$ and $\hat{e}_s$ represent unit vectors in the direction of incident and scattered light, respectively. For the above-mentioned configuration, incident and scattered light polarization unit vectors are given as $\hat{e}_i = [\cos(\theta + \theta_0) \quad \sin(\theta + \theta_0) \quad 0]$ and $\hat{e}_s = [\cos(\theta_0) \quad \sin(\theta_0) \quad 0]$, where $\theta$ is the relative angle between $\hat{e}_i$ and $\hat{e}_s$, whereas $\theta_0$ is an arbitrary angle of scattered light from the X-axis when polarization unit vectors are projected in the XY-plane (*ab*-plane) as shown in the schematic, see Supporting Information Fig. S12. The Raman tensors for the phonon modes with $A_g$ and $B_g$ symmetries are listed in Table I. The angular dependence of the intensity of these modes is given as:

$$I_{A_g} = |b\cos(\theta + \theta_0)\cos\theta_0 + c\sin(\theta + \theta_0)\sin\theta_0|^2, \quad (1)$$

$$I_{B_g} = |f\sin(\theta + 2\theta_0)|^2, \quad (2)$$

without the loss of generality, $\theta_0$ may be taken as zero, resulting in the expression (4) and (5) becomes:

$$I_{A_g} = |b|^2 \cos^2\theta, \qquad (3)$$

$$I_{B_g} = |f|^2 \sin^2\theta. \qquad (4)$$

The polarization dependence of the intensity of some phonon modes cannot be fitted using purely real tensor elements; therefore, a complex form of the Raman tensor should be used to describe the observed polarization dependence [2]. Here, the tensor element of $A_g$ and $B_g$ can be written as: $b = |b|e^{i\phi_b}$ and $c = |c|e^{i\phi_c}$, where $\phi_b$ and $\phi_c$ are the corresponding phases. Incorporating these complex elements, along with the polarization vectors $\hat{e}_i$ and $\hat{e}_s$, and the Raman tensor elements, into eqn (1) yields a modified equation of angular dependence intensities for $A_g$ and $B_g$ given as eqn (4) and (5) (see the main text).

## S2. Temperature-dependent Photoluminescence:

Bulk AgCrP$_2$S$_6$ shows an indirect band gap, however, its flat band structure suggests that it may behave like a direct bandgap material. The photoluminescence (PL) spectra recorded using 532 nm laser displays a peak at ~ 1.32/1.34 eV at temperature ~ 6/300 K. Figure S14(a) represent the temperature evolution of the PL spectra in the temperature range of 6-300 K. The obtained spectrum shows an asymmetric lineshape; therefore, it is fitted using a bi-Gaussian function to extract the temperature-dependent energy (peak position), linewidth, and intensity. The asymmetry observed on the low-energy side of the PL spectrum may have many origins. At low temperatures, the occupation of lower energy states increases, potentially enhancing the contribution of dark excitons or recombination pathways involving localized states, which can leads to asymmetry [3]. Additionally, preferential localization of cations within lattice may induces local dipole moments (stark effect), this can also contribute to asymmetry in PL spectra [4]. Figure S14(c-i) shows the temperature-dependent PL energy of AgCrP$_2$S$_6$ as extracted from our measurements. We observed a redshift up to ~ 40 K, possibly due to the lowering of the conduction band energy by the exchange interaction of the magnetic ions and conduction electrons [5]. Furthermore, at ~ 90 K a sharp change in slope is observed indicating anomaly at this temperature. A kink is observed at ~ 200 K, indicating another anomaly at ~ 200 K. Interestingly, temperature evolution of the PL spectra exhibits an unusual blueshift with increasing temperature, contrary to the typical behaviour observed in most semiconductors (see Fig. S14(b)). This anomalous temperature dependence cannot be accurately described using traditional empirical models like the Varshni [6], Pässler [7], and Bose-Einstein [8], which are commonly used for conventional semiconductors. This discrepancy likely arises from competing effects of electron-phonon renormalization and lattice thermal expansion on the band gap energy [9]. Similar type of behaviour is also reported for some perovskites [9,10]. From linear fitting, the energy obtained at 0 K is represented as $E_0$ = 1.32 eV. The continuous blue-shift in energy above temperature 90 K can be characterized by a temperature coefficient $\alpha = \partial E/\partial T = 0.071$ meVK$^{-1}$.

The temperature-dependent FWHM, denoted as $\Delta(T)$ and determined from the spectral lineshape, is plotted in Fig. S14(c-ii). It is observed that FWHM shows a kink ~ 40 K, after which it starts rapidly increasing from ~ 90 K. We use a linear exciton-phonon coupling model, commonly employed to analyse $\Delta(T)$, given as [11]:

$$\Delta(T) = \Delta_0 + \lambda_{AC}T + \lambda_{LO}N_{LO}(T), \qquad (5)$$

where the $\Delta_0$ term accounts for linewidth broadening due to various factors such as impurities, dislocations, and imperfections. The second term of equation arises from exciton interactions with acoustic phonons. The final term arises from exciton interactions with longitudinal optical (LO) phonon and is proportional to the Bose function, which represents the population of phonons at finite temperature and is given as: $N_{LO}(T) = 1/[e^{(E_{LO}/k_b T)} - 1]$; where $E_{LO}$ is the energy of LO phonons and $k_b$ is the Boltzmann constant. The coefficients $\lambda_{AC}$ and $\lambda_{LO}$ represent coupling strength of the exciton-acoustic/LO phonons, respectively. Exciton-LO phonon interactions are understood through the Fröhlich mechanism, while exciton-acoustic phonon interactions are described by the deformation potential. From the fitting eqn (5), the parameters obtained as follows: $\Delta_0 = 91.16 \pm 2.39$ meV, $\lambda_{AC} = 203.76 \pm 18.60$ μeVK$^{-1}$, $\lambda_{LO} = 3.98 \pm 2.46$ eV, $E_{LO} = 123.46 \pm 17.69$ meV. The exciton-acoustic phonon coupling strength is orders of magnitude smaller than the exciton-optical phonon coupling strength, indicating that optical phonons contribute significantly more to linewidth broadening than acoustic phonons. This is consistent with the observed large LO phonon energy. The temperature-dependent (PL) intensity, shown in Fig. S14(c-iii), decreases drastically from 6 K to 90 K. This intensity decrease is primarily attributed to thermal quenching. At low temperatures, the luminescence internal quantum efficiency is high because non-radiative recombination centres are minimized. As the temperature increases, these centres become thermally activated, capturing more photogenerated carriers. This, in turn, reduces the number of carriers available for radiative recombination, resulting in the observed decrease in the PL intensity [12]. Above ~ 200 K, the intensity then gradually increases with rising temperature, reason associated with this anomalous increase of intensity is unknown; however, we should stress that the anomalous temperature dependent PL of AgCrP$_2$S$_6$ demands further theoretical and experimental investigations.

**Figures:**

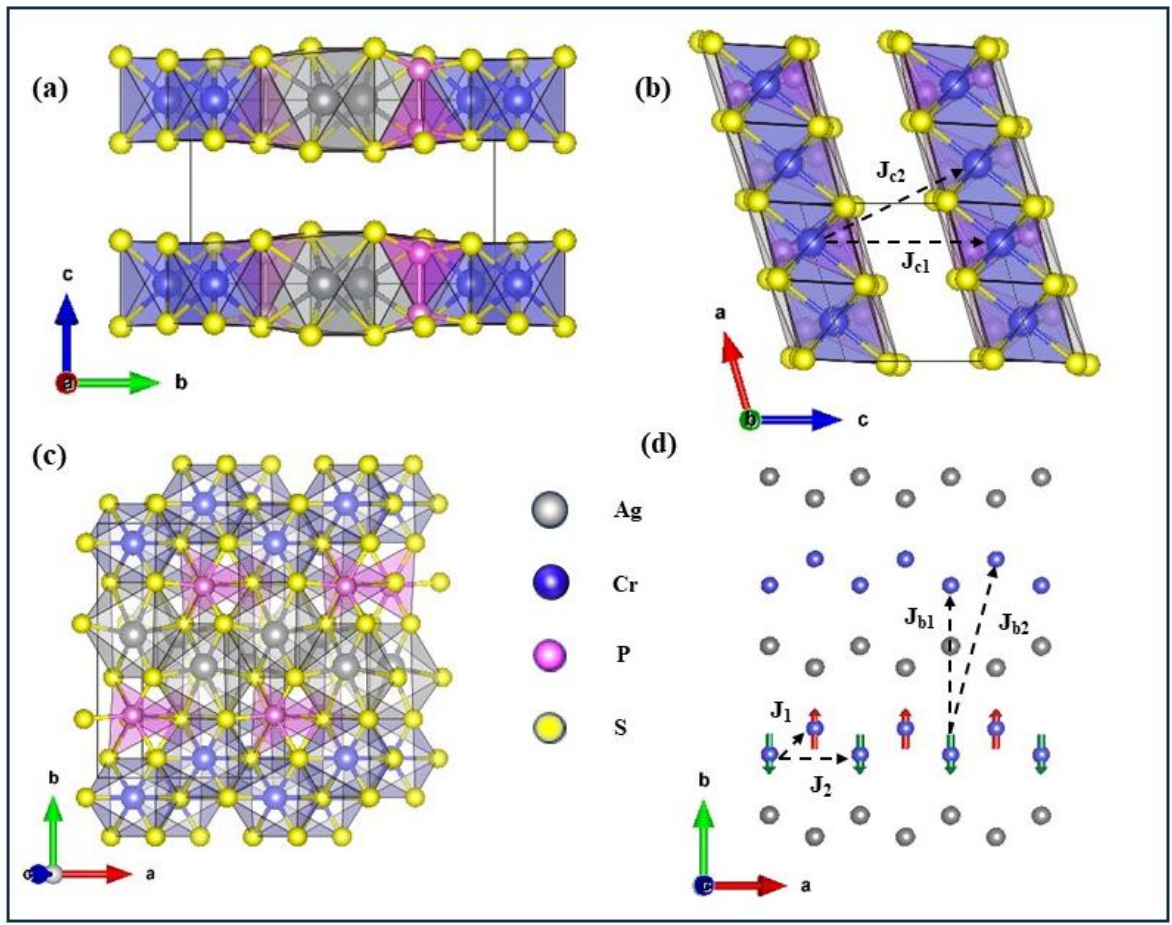

**Figure S1. (a)-(c)** Shows the crystal structure of AgCrP$_2$S$_6$ in the *bc*, *ac*, and *ab* plane, respectively. (d) Zigzag chains of magnetic Cr$^{3+}$ ions (spin up and down shown by arrows) and nonmagnetic Ag$^+$ ions.

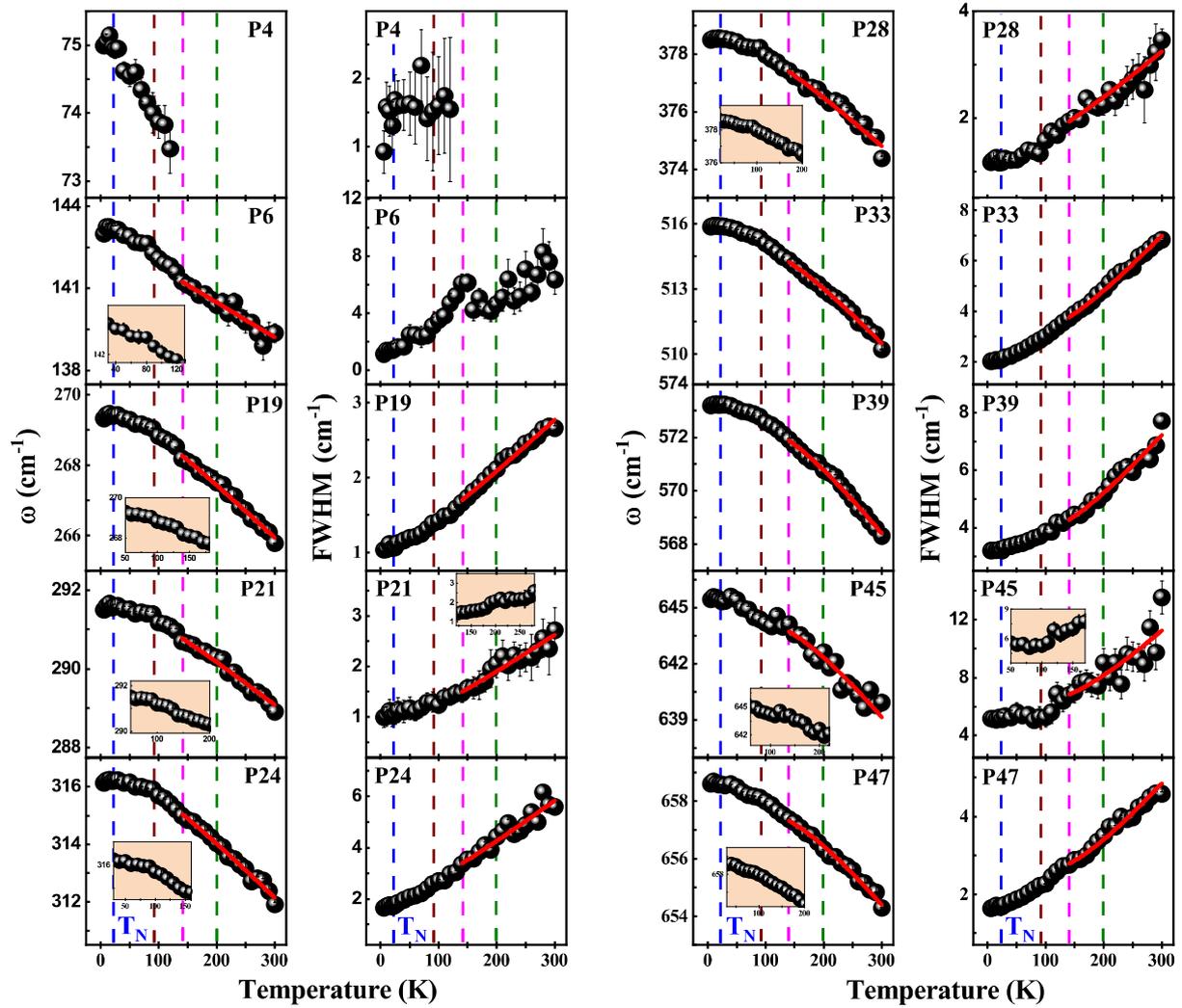

**Figure S2.** Temperature evolution of mode frequency and linewidth (FWHM) of phonon modes P4, P6, P19, P21, P24, P28, P33, P39, P45 and P47. The solid red line represents a three-phonon fitting in the temperature range of 140 to 300 K, whereas the dashed blue line indicates the antiferromagnetic transition at $T_N \sim 20$ K. The dashed brown, pink, and green lines reflect temperature anomalies at $\sim 90$ K, $\sim 140$ K, and $\sim 200$ K, respectively. Inset shows the magnified image of corresponding anomalies in frequency and FWHM.

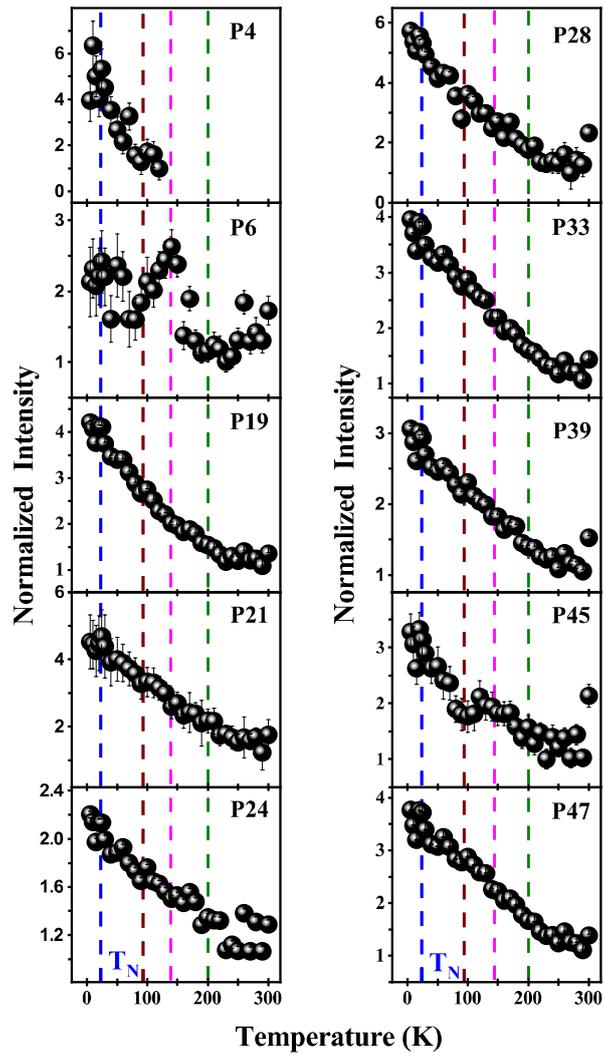

**Figure S3.** Shows temperature dependence of normalised intensity of modes P4, P6, P19, P21, P24, P28, P33, P39, P45 and P47.

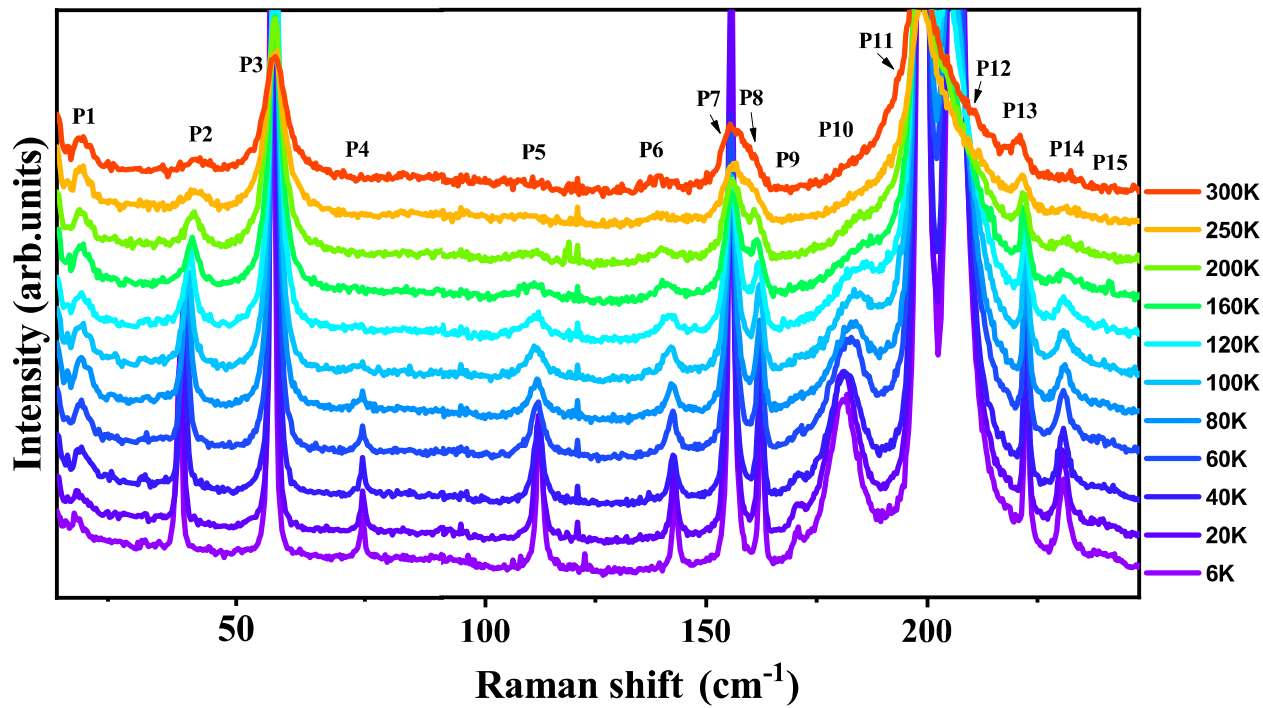

**Figure S4.** Shows Temperature evolution of the Raman spectrum of $AgCrP_2S_6$ in the frequency range of 15-250 cm$^{-1}$. Mode P4, P5, P9, and P15 disappeared at higher temperature.

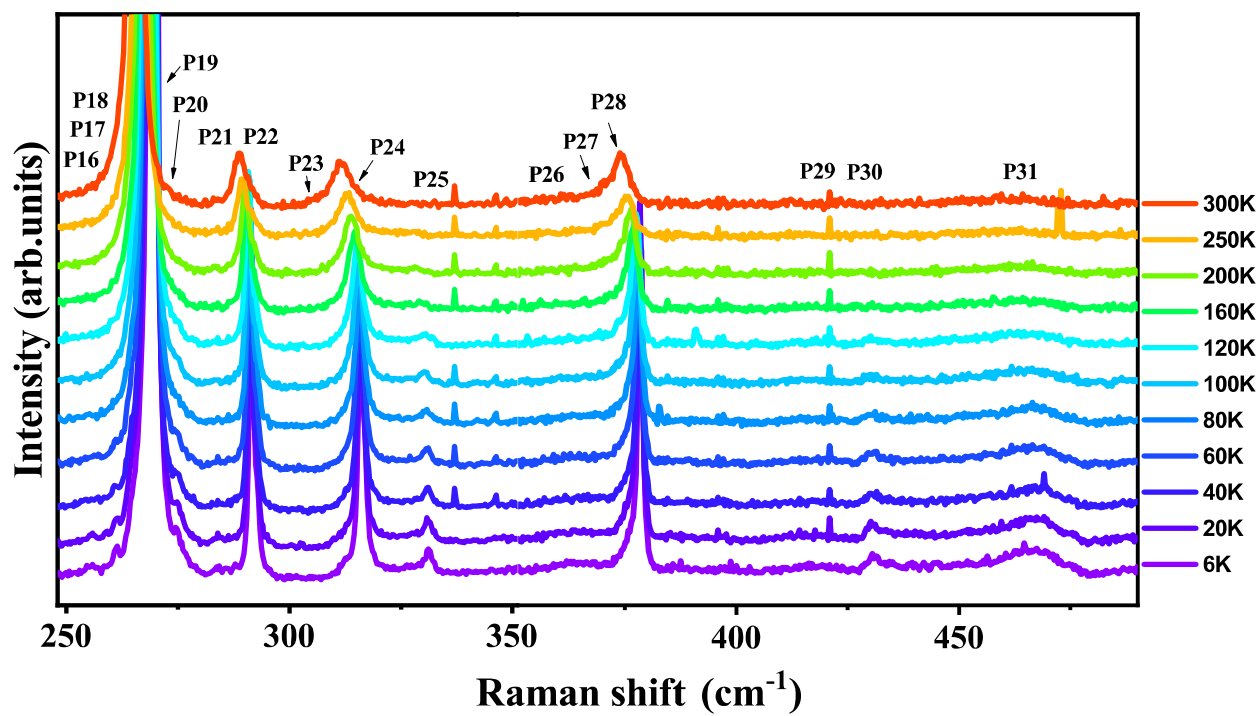

**Figure S5.** Shows Temperature evolution of the Raman spectrum of $AgCrP_2S_6$ in the frequency range of 250-490 cm$^{-1}$. Mode P16, P17, P20, P23, P25, P29, P30, and 31 disappeared at higher temperature.

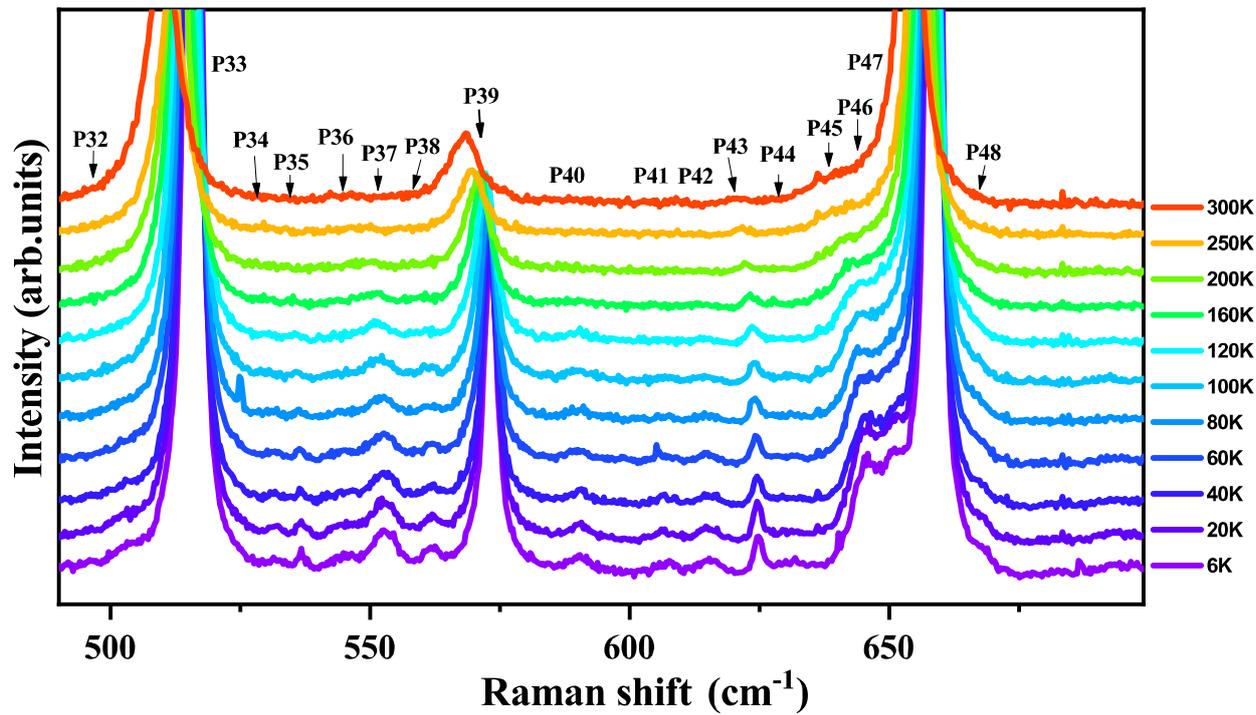

**Figure S6.** Shows Temperature evolution of the Raman spectrum of $AgCrP_2S_6$ in the frequency range of 490-700 cm$^{-1}$. Mode P32, P34, P35, P36, P40, P41, P42, P44, and P48 disappeared at higher temperature.

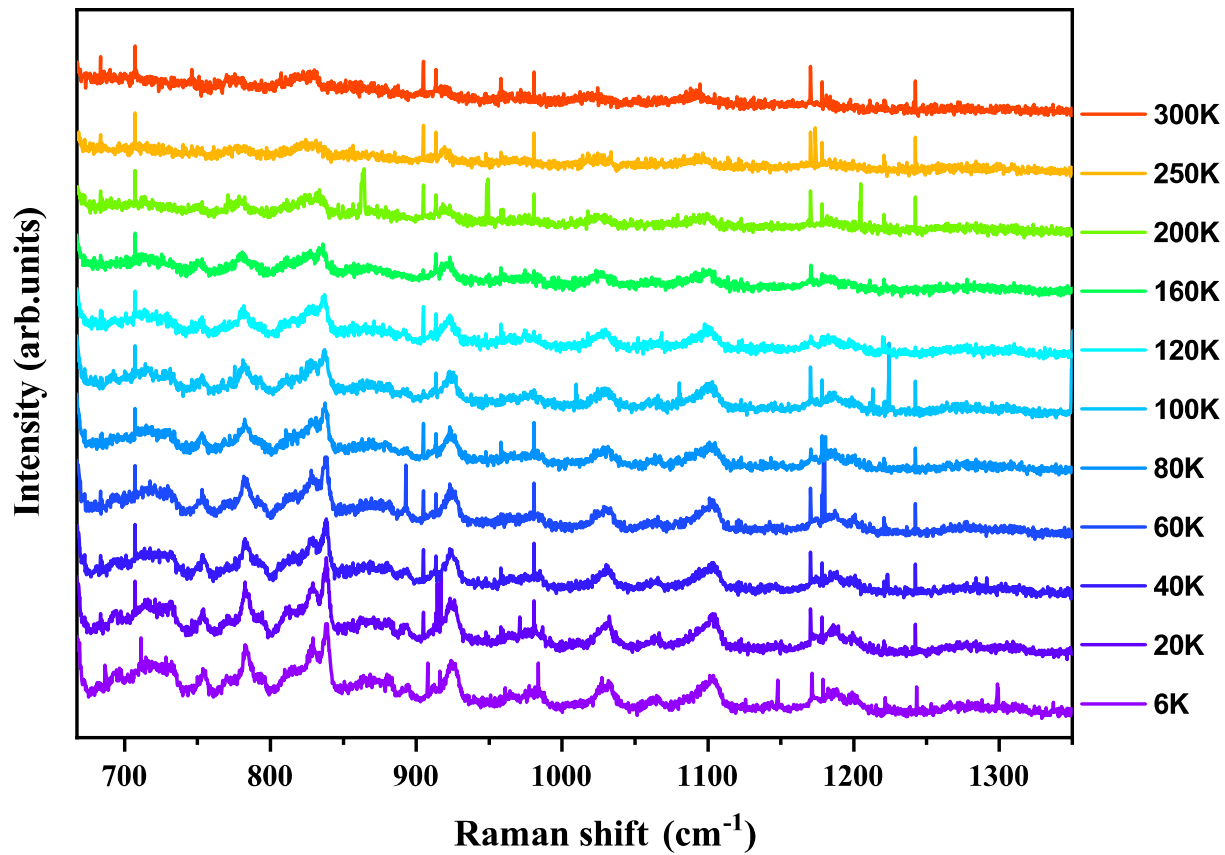

**Figure S7.** Temperature evolution of higher order modes in the frequency range of 670 cm$^{-1}$ to 1350 cm$^{-1}$.

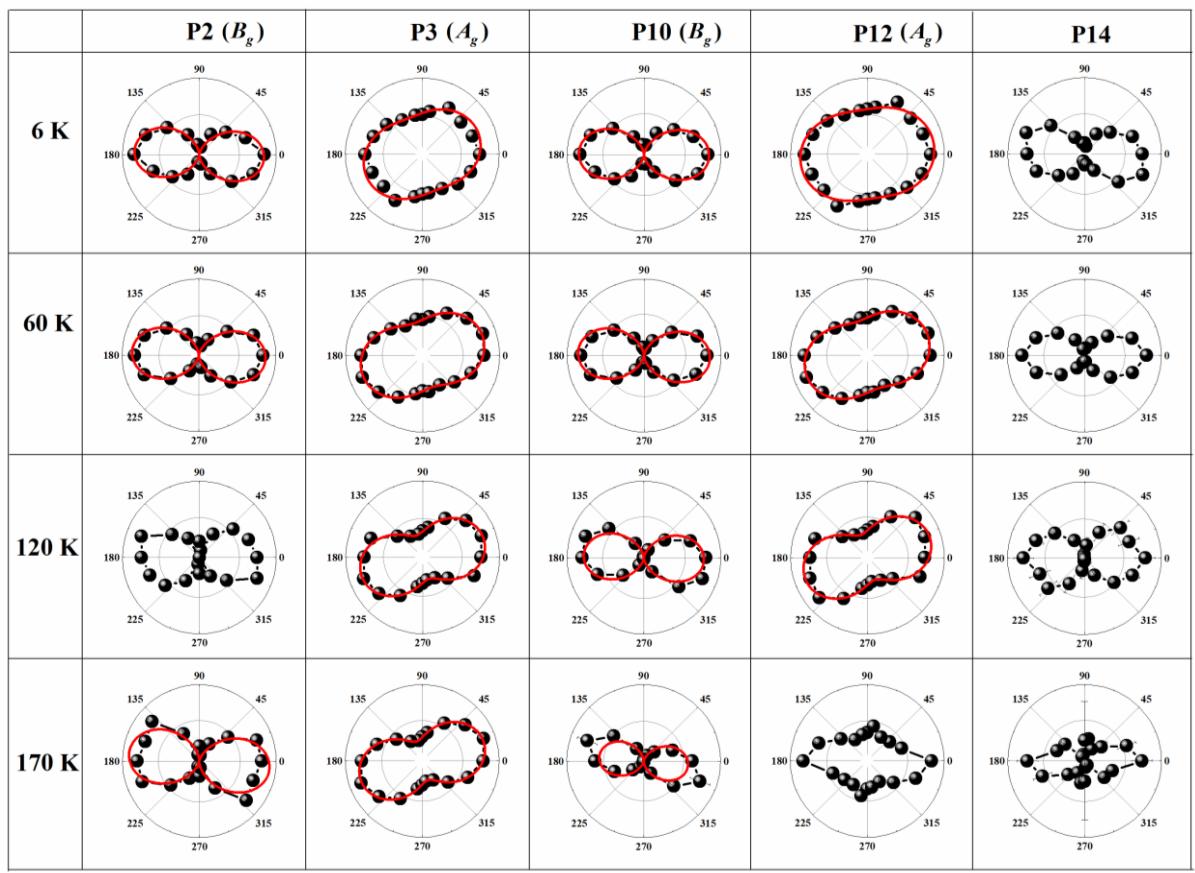

**Figure S8.** Polarization-dependent intensity of the phonon modes P2, P3, P10, P12, and P14 measured at temperatures of 6 K, 60 K, 120 K, and 170 K, respectively; with varying incident light polarization. The solid red line at the bottom indicates the direction of the scattered light polarization.

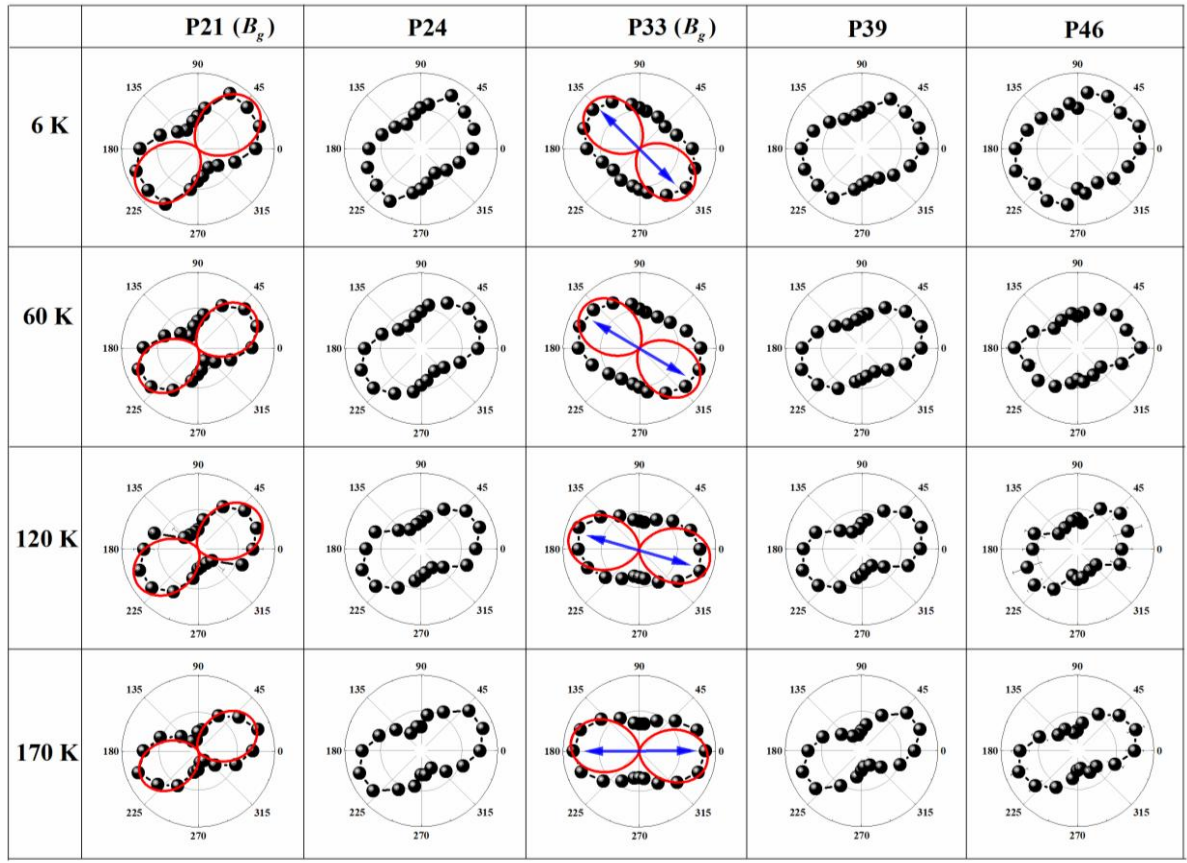

**Figure S9.** Polarization-dependent intensity of the phonon modes P21, P24, P33, P39, and P46 measured at temperatures of 6 K, 60 K, 120 K, and 170 K, respectively; with varying incident light polarization. A blue arrow represents the rotation of the axis, showing the maxima as a function of temperature. The solid red line at the bottom indicates the direction of the scattered light polarization.

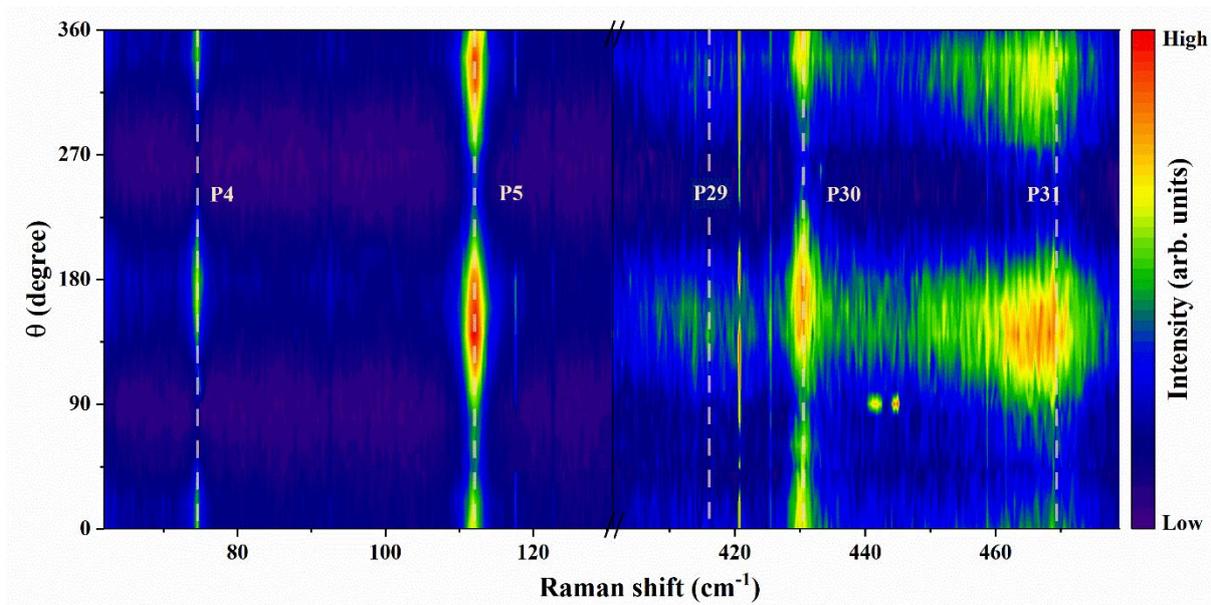

**Figure S10.** 2D colour contour map of polarization-dependent intensity of the phonon modes P4, P5, P29, P30, and P31 measured at 6 K.

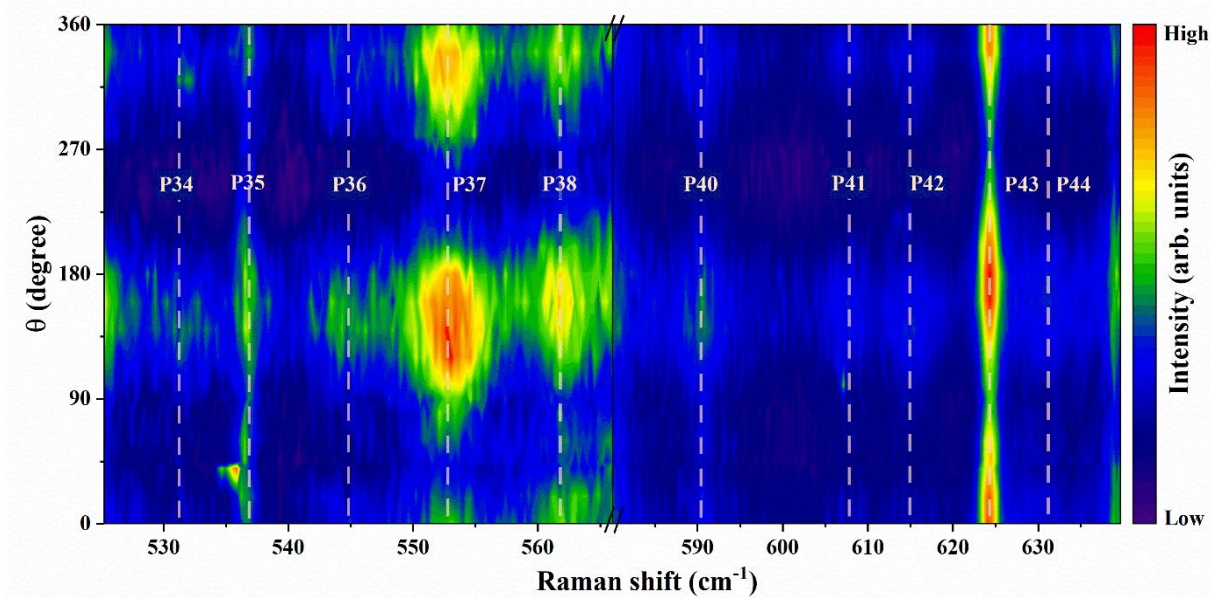

**Figure S11.** 2D colour contour map of polarization-dependent intensity of the phonon modes P34, P35, P36, P37, P38, P40, P41, P42, P43 and P44 measured at 6 K.

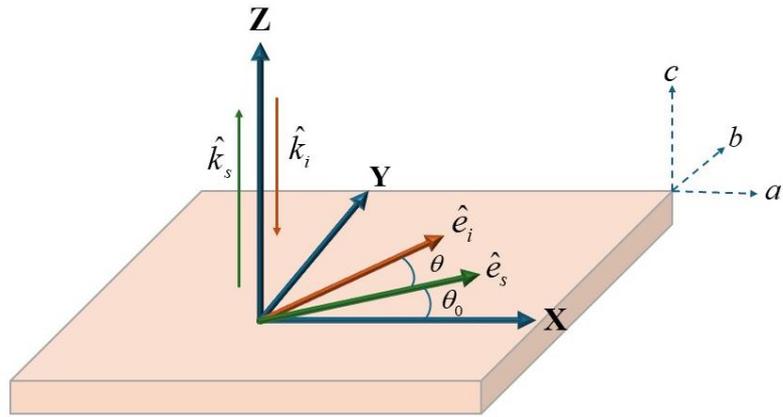

**Figure S12.** Schematic showing plane projection of the polarization direction of the incident ($\hat{e}_i$) and scattered light ($\hat{e}_s$) making angle $\theta+\theta_0$ and $\theta$ with X-axis, respectively. $\hat{k}_i$ and $\hat{k}_s$ represents the direction of incident and scattered beam.

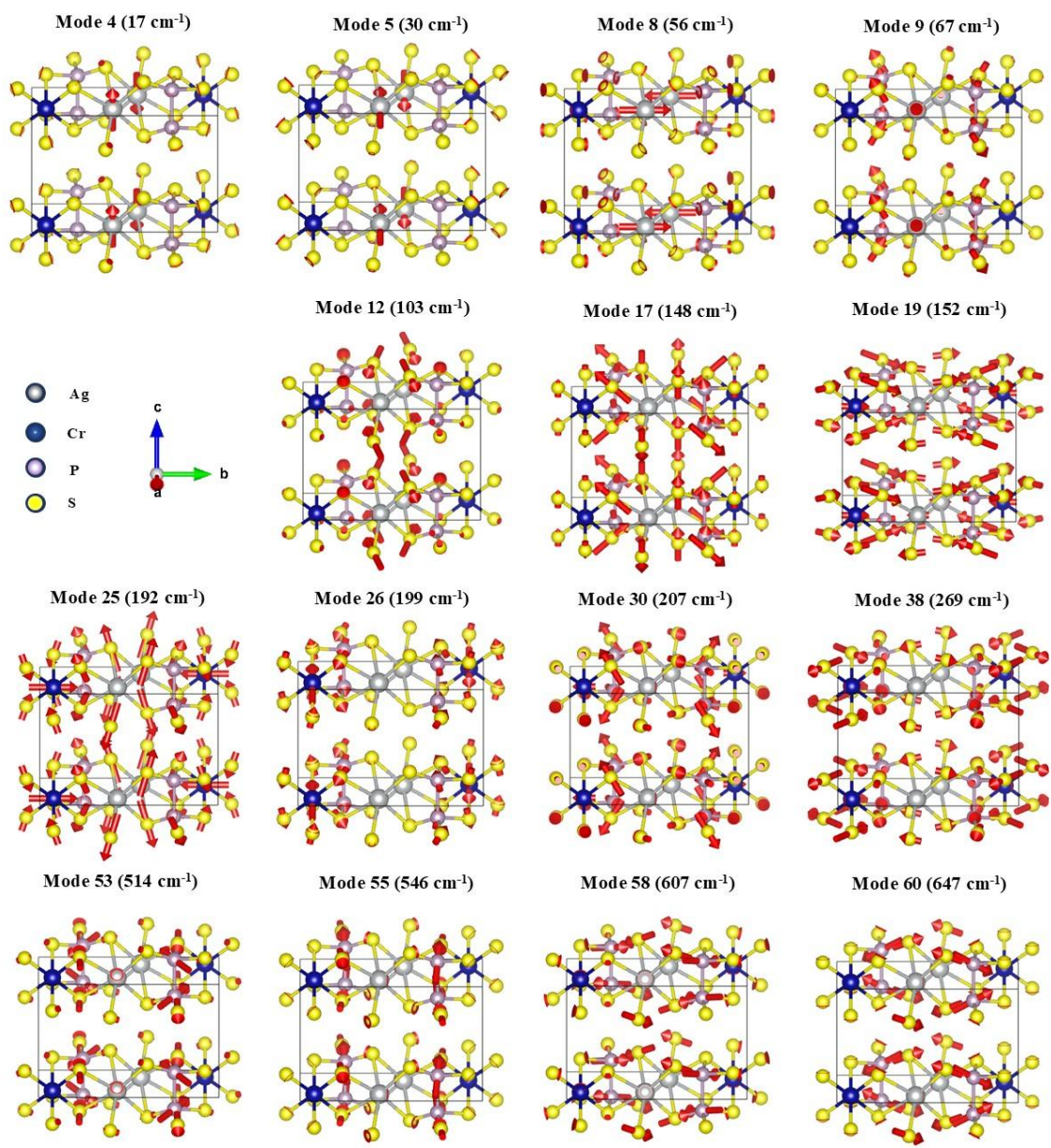

**Figure S13.** Mode visualization of some of the Raman active modes at Γ- point. Grey, blue, violet, and yellow spheres represent Ag, Cr, P, and S atoms, respectively. Red arrows on the atoms represent the direction of motion of the atoms and the size of the arrows represent the relative magnitude of vibration.

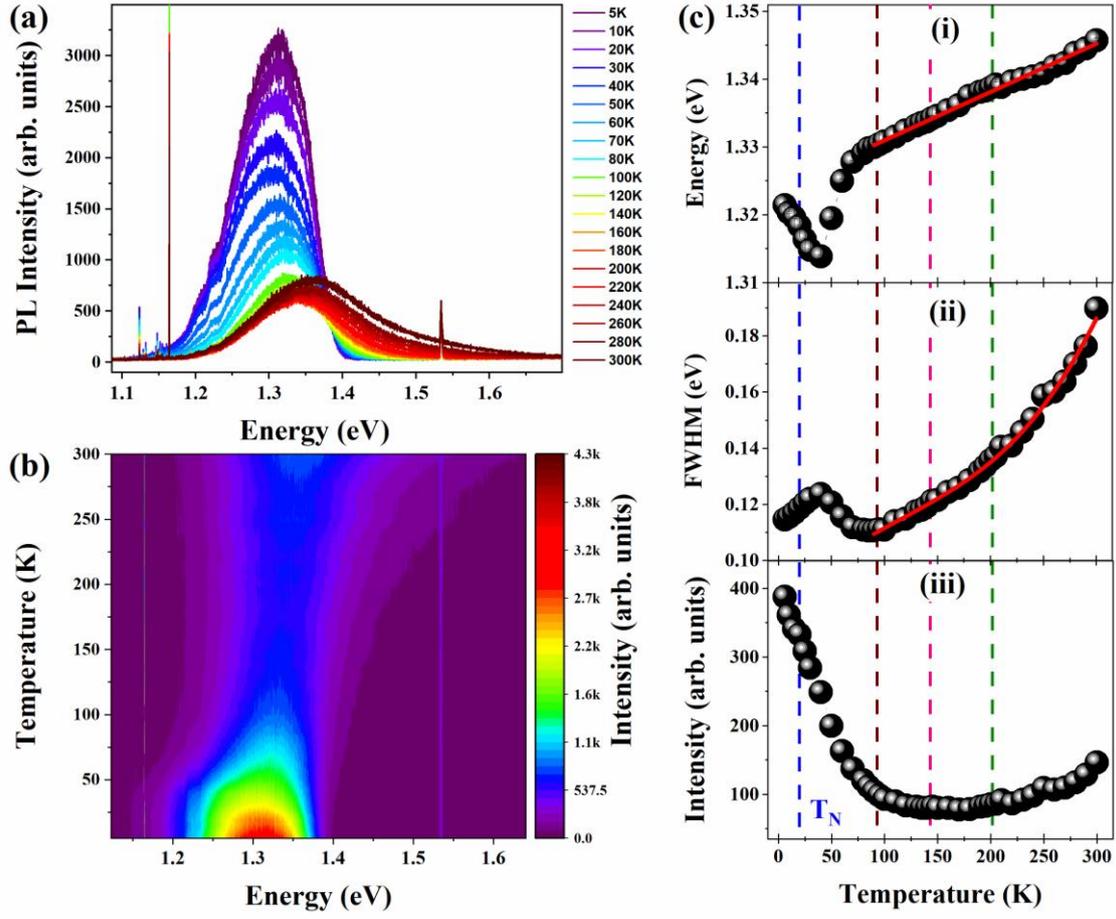

**Figure S14. (a)** Shows temperature evolution of PL. **(b)** Shows a colour contour plot of PL intensity in temperature-energy plane. **(c-i)** Temperature dependence of energy. **(c-ii)** Temperature dependence of FWHM. The red line shows the fitted curve as described in the text. **(c-iii)** Temperature dependence of PL intensity. The dashed blue line indicates the antiferromagnetic transition at $T_N \sim 20$ K. The dashed brown, pink, and green lines reflect temperature anomalies at $\sim 90$ K, $\sim 140$ K, and $\sim 200$ K, respectively.

**Tables:**

**Table S1.** List of the experimentally observed modes with their frequencies at 6 K and 300 K. The calculated phonon frequencies on the basis of first-principles calculations are denoted by $\omega_{DFT}$. Units are in cm$^{-1}$.

| Modes | $\omega$ (6 K) | $\omega$ (300 K) | Symmetry assignment | $\omega_{DFT}$ |
|---|---|---|---|---|
| P1  | 19.0 ± 0.10 | 19.9 ± 0.09 | $B_g$ | 17 |
| P2  | 38.8 ± 0.03 | 42.2 ± 0.18 | $B_g$ | 30 |
| P3  | 56.9 ± 0.01 | 57.5 ± 0.02 | $A_g$ | 56 |
| P4  | 74.9 ± 0.09 | - | $B_g$ | 67 |
| P5  | 112.3 ± 0.03 | - | $A_g$ | 103 |
| P6  | 142.9 ± 0.11 | 139.4 ± 0.34 | $B_g$ | 148 |
| P7  | 155.7 ± 0.01 | 156.4 ± 0.13 | $A_g$ | 152 |
| P8  | 162.5 ± 0.04 | 159.5 ± 0.21 | $B_g$ | 156 |
| P9  | 170.7 ± 0.41 | - | $A_g$ | 169 |
| P10 | 181.0 ± 0.15 | 189.9 ± 0.65 | $B_g$, $A_g$ | 178, 192 |
| P11 | 199.0 ± 0.00 | 198.6 ± 0.05 | $B_g$ | 199 |
| P12 | 206.5 ± 0.01 | 207.1 ± 0.43 | $A_g$ | 207 |
| P13 | 222.5 ± 0.04 | 221.1 ± 0.11 | $B_g$ | 213 |
| P14 | 230.9 ± 0.18 | 232.6 ± 0.45 | - | - |
| P15 | 241.7 ± 4.05 | - | - | - |
| P16 | 254.9 ± 3.81 | - | - | - |
| P17 | 261.2 ± 0.88 | - | - | - |
| P18 | 265.1 ± 0.85 | 263.0 ± 0.26 | $B_g$ | 266 |
| P19 | 269.3 ± 0.00 | 265.8 ± 0.01 | $A_g$ | 269 |
| P20 | 274.9 ± 0.72 | - | $A_g$, $B_g$ | 278, 279 |
| P21 | 291.4 ± 0.06 | 288.9 ± 0.17 | $B_g$ | 293 |
| P22 | 292.8 ± 0.36 | 291.1 ± 0.78 | - | - |
| P23 | 312.1 ± 0.26 | - | - | - |
| P24 | 316.1 ± 0.00 | 311.9 ± 0.06 | - | - |
| P25 | 330.7 ± 0.18 | - | $A_g$ | 346 |
| P26 | 362.2 ± 0.94 | 359.8 ± 1.82 | - | - |
| P27 | 376.1 ± 0.18 | 370.6 ± 0.42 | - | - |
| P28 | 378.5 ± 0.00 | 374.4 ± 0.07 | - | - |
| P29 | 416.7 ± 0.49 | - | $B_g$ | 407 |
| P30 | 431.0 ± 0.11 | - | - | - |
| P31 | 469.5 ± 0.33 | - | $A_g$ | 465 |
| P32 | 503.0 ± 0.91 | - | $A_g$ | 486 |
| P33 | 515.8 ± 0.00 | 510.2 ± 0.01 | $B_g$ | 514 |
| P34 | 531.6 ± 0.73 | - | - | - |
| P35 | 536.6 ± 0.24 | - | - | - |
| P36 | 545.0 ± 0.75 | - | $A_g$ | 546 |
| P37 | 552.7 ± 0.25 | 545.4 ± 0.89 | - | - |
| P38 | 561.9 ± 0.41 | 554.0 ± 1.78 | - | - |
| P39 | 573.1 ± 0.02 | 568.3 ± 0.06 | - | - |
| P40 | 590.3 ± 0.59 | - | - | - |
| P41 | 607.2 ± 0.78 | - | $B_g$ | 607 |
| P42 | 615.2 ± 0.56 | - | - | - |
| P43 | 624.8 ± 0.13 | 620.4 ± 0.41 | - | - |
| P44 | 631.5 ± 0.93 | - | - | - |
| P45 | 645.4 ± 0.14 | 639.9 ± 0.41 | - | - |
| P46 | 650.7 ± 0.16 | 650.5 ± 0.44 | - | - |
| P47 | 658.6 ± 0.00 | 654.3 ± 0.02 | $A_g$ | 647 |
| P48 | 668.5 ± 4.19 | - | - | - |

**Table S2.** Frequencies and symmetry labels of all 60 modes based on first-principles calculations of phonon frequencies at $\Gamma$ - point for AgCrP$_2$S$_6$.

| Mode No. | Frequency | Symmetry | Mode No. | Frequency | Symmetry |
|---|---|---|---|---|---|
| 1 | 0 | $B_u$ | 31 | 210 | $B_u$ |
| 2 | 0 | $A_u$ | 32 | 213 | $B_g$ |
| 3 | 0 | $B_u$ | 33 | 223 | $B_u$ |
| 4 | 17 | $B_g$ | 34 | 227 | $A_u$ |
| 5 | 30 | $B_g$ | 35 | 245 | $A_u$ |
| 6 | 37 | $B_u$ | 36 | 250 | $B_u$ |
| 7 | 43 | $B_u$ | 37 | 266 | $B_g$ |
| 8 | 56 | $A_g$ | 38 | 269 | $A_g$ |
| 9 | 67 | $B_g$ | 39 | 278 | $A_g$ |
| 10 | 78 | $A_u$ | 40 | 279 | $B_g$ |
| 11 | 92 | $B_g$ | 41 | 282 | $B_u$ |
| 12 | 103 | $A_g$ | 42 | 293 | $B_g$ |
| 13 | 111 | $B_u$ | 43 | 295 | $A_u$ |
| 14 | 113 | $A_u$ | 44 | 317 | $B_u$ |
| 15 | 125 | $B_u$ | 45 | 346 | $A_u$ |

| 16 | 135 | $B_g$ | 46 | 346 | $A_g$ |
|---|---|---|---|---|---|
| 17 | 148 | $B_g$ | 47 | 399 | $B_u$ |
| 18 | 150 | $B_u$ | 48 | 407 | $B_g$ |
| 19 | 152 | $A_g$ | 49 | 427 | $A_u$ |
| 20 | 156 | $B_g$ | 50 | 465 | $A_g$ |
| 21 | 169 | $A_g$ | 51 | 486 | $A_g$ |
| 22 | 174 | $A_u$ | 52 | 487 | $A_u$ |
| 23 | 178 | $B_g$ | 53 | 514 | $B_g$ |
| 24 | 184 | $A_u$ | 54 | 517 | $B_u$ |
| 25 | 192 | $A_g$ | 55 | 546 | $A_g$ |
| 26 | 199 | $B_g$ | 56 | 550 | $A_u$ |
| 27 | 201 | $A_g$ | 57 | 606 | $B_u$ |
| 28 | 201 | $B_u$ | 58 | 607 | $B_g$ |
| 29 | 206 | $A_u$ | 59 | 620 | $A_u$ |
| 30 | 207 | $A_g$ | 60 | 647 | $A_g$ |

**Table S3.** List of fitting parameters for modes shown in Fig. 4 and Fig. S2 using three-phonon anharmonic model in the temperature range 140 K to 300 K. All units are in cm$^{-1}$.

| Mode | $\omega_0$ | A | $\Gamma_0$ | C |
|---|---|---|---|---|
| P2  | 40.3 ± 0.09  | 0.09 ± 0.01   | -            | -            |
| P6  | 143.1 ± 0.16 | -0.67 ± 0.04  | -            | -            |
| P8  | 163.7 ± 0.16 | -0.74 ± 0.05  | -            | -            |
| P10 | 180.0 ± 0.47 | 1.62 ± 0.15   | -            | -            |
| P13 | 223.2 ± 0.10 | -0.48 ± 0.04  | 0.06 ± 0.28  | 0.86 ± 0.11  |
| P19 | 270.8 ± 0.09 | -1.53 ± 0.04  | 0.52 ± 0.05  | 0.71 ± 0.02  |
| P21 | 292.7 ± 0.09 | -1.22 ± 0.05  | 0.22 ± 0.12  | 0.81 ± 0.06  |
| P24 | 318.5 ± 0.11 | -2.31 ± 0.06  | 0.49 ± 0.26  | 1.95 ± 0.14  |
| P28 | 380.8 ± 0.20 | -2.56 ± 0.13  | 0.21 ± 0.19  | 1.29 ± 0.12  |
| P33 | 520.8 ± 0.15 | -5.73 ± 0.11  | -1.92 ± 0.14 | 4.94 ± 0.10  |
| P39 | 578.8 ± 0.16 | -6.25 ± 0.12  | -1.61 ± 0.33 | 5.31 ± 0.25  |
| P45 | 654.5 ± 0.97 | -10.11 ± 0.79 | -3.67 ± 1.80 | 9.79 ± 1.45  |
| P47 | 664.4 ± 0.21 | -6.64 ± 0.18  | -2.23 ± 0.24 | 4.69 ± 0.19  |